\pdfoutput=1

\documentclass[final,twocolumn]{raa_twocolumn}           

\usepackage{graphicx,times}
\usepackage{natbib}
\usepackage{amssymb,amsmath}
\usepackage{multirow}

\bibpunct{(}{)}{;}{a}{}{,}

\usepackage[pagebackref=true,
            citecolor=blue,
            colorlinks=true,
            linkcolor=blue,
            filecolor=blue,  
            urlcolor=blue,      
            final=true,
            ]{hyperref}

\usepackage{ulem}

\usepackage{CJK}

\begin{document}
\begin{CJK*}{UTF8}{gbsn}

\title{The ALMA-QUARKS survey: 
-- I. Survey description and data reduction
}

 \volnopage{ {\bf 20XX} Vol.\ {\bf X} No. {\bf XX}, 000--000}
   \setcounter{page}{1}

\author{
Xunchuan Liu (刘训川) \inst{1},
Tie Liu \inst{1},
Lei Zhu \inst{2},
Guido Garay \inst{3},
Hong-Li Liu \inst{4},
Paul Goldsmith \inst{5},
Neal Evans \inst{6},
Kee-Tae Kim \inst{7},
Sheng-Yuan Liu \inst{8},
Fengwei Xu \inst{9,10},
Xing Lu \inst{1},
Anandmayee Tej \inst{11},
Xiaofeng Mai \inst{1},
Leonardo Bronfman \inst{3},
Shanghuo Li \inst{12},
Diego Mardones \inst{3},
Amelia Stutz \inst{13},
Ken'ichi Tatematsu \inst{14},
Ke Wang \inst{9},
Qizhou Zhang \inst{15},
Sheng-Li Qin \inst{4},
Jianwen Zhou \inst{16},
Qiuyi Luo \inst{1,17},
Siju Zhang \inst{9},
Yu Cheng \inst{18},
Jinhua He \inst{19,2,3},
Qilao Gu \inst{1},
Ziyang Li \inst{1,4},
Zhenying Zhang \inst{1,4},
Suinan Zhang \inst{1},
Anindya Saha \inst{11},
Lokesh Dewangan \inst{20},
Patricio Sanhueza \inst{18,21},
Zhiqiang Shen \inst{1},
}

\institute{ 
Shanghai Astronomical Observatory, Chinese Academy of Sciences, Shanghai 200030, PR China; {\it liuxunchuan@shao.ac.cn, liutie@shao.ac.cn}\\
\and
Chinese Academy of Sciences South America Center for Astronomy, National Astronomical Observatories, Chinese Academy of Sciences, Beijing, 100101, PR China; {\it lzhupku@gmail.com}\\
\and
Departamento de Astronom\'ia, Universidad de Chile, Las Condes, 7591245 Santiago, Chile; {\it guido@das.uchile.cl}\\
\and
School of Physics and Astronomy, Yunnan University, Kunming, 650091, PR China; {\it hongliliu2012@gmail.com}\\
\and
Jet Propulsion Laboratory, California Institute of Technology, 4800 Oak Grove Drive, Pasadena CA 91109, USA\\
\and
Department of Astronomy, The University of Texas at Austin, Texas 78712-1205, USA\\
\and
Korea Astronomy and Space Science Institute, 776 Daedeokdae-ro, Yuseong-gu, Daejeon 34055, Republic of Korea\\
\and
Institute of Astronomy and Astrophysics, Academia Sinica, Roosevelt Road, Taipei 10617, Taiwan (R.O.C)\\
\and
Kavli Institute for Astronomy and Astrophysics, Peking University, Haidian District, Beijing 100871, PR China\\
\and
Department of Astronomy, School of Physics, Peking University, Beijing 100871, PR China\\
\and
Indian Institute of Space Science and Technology, Thiruvananthapuram 695 547, Kerala, India\\
\and
Max Planck Institute for Astronomy, K\"onigstuhl 17, D-69117 Heidelberg, Germany\\
\and
Departamento de Astronom\'ia, Universidad de Concepci\'on, Casilla 160-C, Concepci\'on, Chile\\
\and
Nobeyama Radio Observatory, National Astronomical Observatory of Japan, National Institutes of Natural Sciences, 462-2 Nobeyama, Minamimaki, Minamisaku, Nagano 384-1305, Japan\\
\and
Center for Astrophysics $|$ Harvard \& Smithsonian, 60 Garden Street, Cambridge, MA 02138, USA\\
\and
Max-Planck-Institut f\"ur Radioastronomie, Auf dem H\"ugel 69, 53125 Bonn, Germany\\
\and
School of Astronomy and Space Sciences, University of Chinese Academy of Sciences, Beijing 100049, PR China\\
\and
National Astronomical Observatory of Japan, 2-21-1 Osawa, Mitaka, Tokyo, 181-8588, Japan\\
\and
Yunnan Observatories, Chinese Academy of Sciences, Kunming, 650216, Yunnan, PR China\\
\and
Physical Research Laboratory, Navrangpura, Ahmedabad-380 009, India\\
\and
GUAS Astronomical Science Program, SOKENDAI, 2-21-1 Osawa, Mitaka, Tokyo 181-8588, Japan\\
\vs \no
   {\small Received 20XX Month Day; accepted 20XX Month Day}
}

\abstract{
This paper presents an overview of the QUARKS survey, 
which stands for `Querying Underlying mechanisms of massive star formation with ALMA-Resolved gas
Kinematics and Structures'. The QUARKS survey 
is observing 139 massive  clumps covered by 156 pointings at ALMA Band 6  ($\lambda\sim$ 1.3 mm).  
In conjunction with data obtained from the ALMA-ATOMS survey at Band 3 ($\lambda\sim$ 3 mm), QUARKS aims to  carry out an unbiased statistical
investigation of massive star formation process within  protoclusters down to a scale of 1000 au.
This overview paper describes the observations and data reduction of the QUARKS survey, 
and gives a first look at an exemplar source, the mini-starburst Sgr B2(M).
The wide-bandwidth (7.5 GHz) and high-angular-resolution ($\sim 0.3\arcsec$) observations of the QUARKS survey 
allow to resolve much more compact cores than could be done by the ATOMS survey, and to detect
previously unrevealed fainter filamentary structures. 
The spectral windows cover transitions of species including CO, SO, N$_2$D$^+$, SiO, H$_{30}\alpha$, H$_2$CO, CH$_3$CN and many other
complex organic molecules, tracing gas components with different temperatures and spatial extents.
QUARKS aims to deepen our understanding of several scientific topics of massive star formation, such as the mass transport within protoclusters by
(hub-)filamentary structures, the existence of massive starless cores, the physical and chemical properties
of dense cores within protoclusters, and the feedback from already formed high-mass young protostars.
}

   \authorrunning{Xunchuan Liu, Tie Liu, Lei Zhu, et al.}            
   \titlerunning{QUARKS survey description}  
   \maketitle

%
\section{Introduction}           
\label{sect:intro}

As the principal sources of heavy elements and UV radiation, high-mass stars ($M>8$ $M_{\sun}$) play a
major role in the evolution of galaxies. However, 
the formation mechanism of high-mass stars is still under intense debate
\citep[e.g.,][]{2007ARA&A..45..481Z,2023ASPC..534..233P}.
A variety of models have been proposed, 
contending with each other or focusing on  different aspects of high-mass star formation.
Early models disagreed about how a massive protostar obtains its initial mass and 
subsequently accretes additional mass.
The competitive accretion scenario \citep[e.g.,][]{2004MNRAS.349..735B,2006MNRAS.370..488B}  assumed initial 
fragmentation of a molecular cloud into low-mass cores each with 
mass approximately equal to the thermal Jeans mass. In contrast, the 
``turbulent core accretion model'' \citep[e.g.,][]{2003ApJ...585..850M}
suggested that high-mass stars form directly from turbulent massive gas cores. 
These idealized  models  depict  protostars forming and accreting
from their natal gas reservoirs which are treated as unstructured and obeying isotropic dynamics. Therefore, models that explained the presence of star forming regions 
in/on hierarchical structures, or more specifically filamentary structures, were put forward
\citep[e.g., global hierarchical collapse model, inertial-inflow model;][]{2008ApJ...674..316H,2009ApJ...707.1023V,2020ApJ...900...82P}.
These models were partly   
supported by  the observations of far-infrared dust continuum by the {\it Herschel} space observatory and other ground-based single-dishes,
which demonstrated that filaments are ubiquitous and closely correlated with star formation in nearby molecular clouds as well as massive clumps in the Galactic Plane \citep[e.g.,][]{2010A&A...518L.102A,2015A&A...584A..91K,2016A&A...591A...5L,2018ApJS..234...28L,2020A&A...642A..87K,2020MNRAS.492.5420S,2021ApJ...912..148L,2023ASPC..534..153H,2023A&A...675A.119G}.

\begin{figure*}[!thb]
\centering
\includegraphics[width=0.999\linewidth]{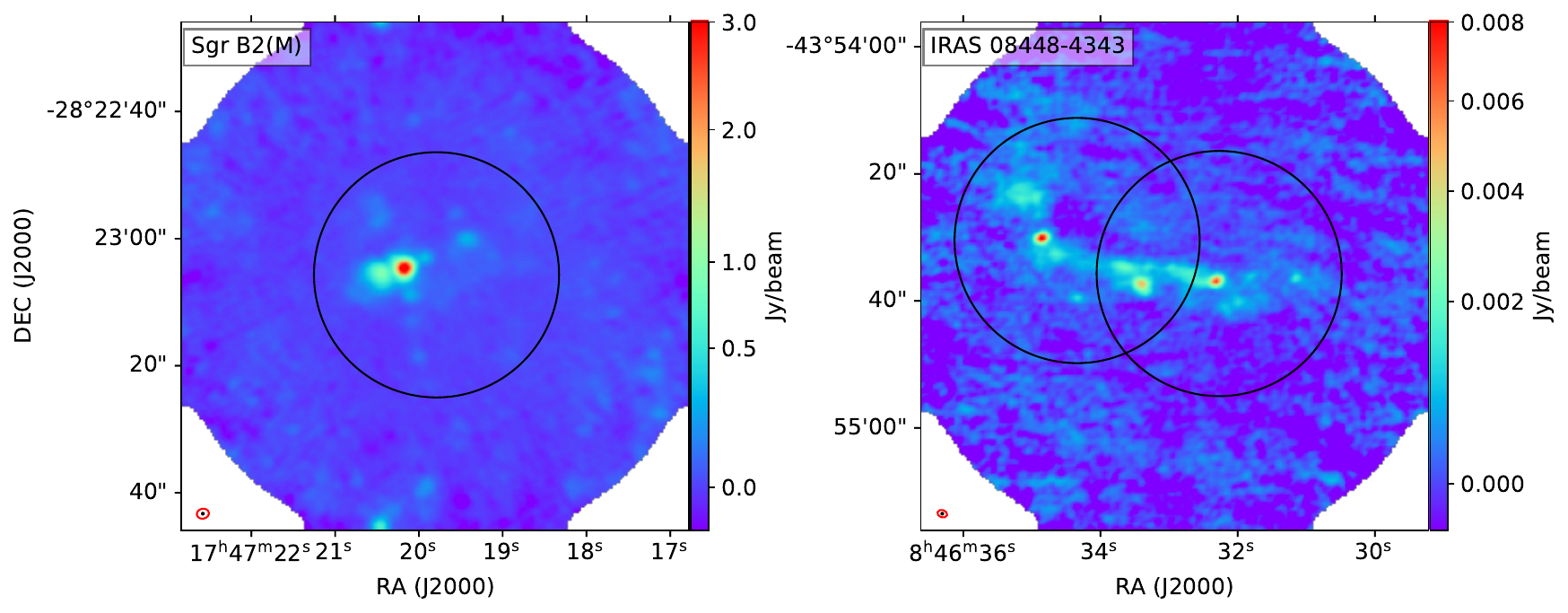}
\caption{Example of the targets of the QUARKS survey. In each panel, the background is the Band 3 continuum of the ATOMS survey.
The black circle is primary beam of the ALMA 12-m array  in Band 6.
The red ellipse and the black dot in the lower-left corner show the angular resolutions of the ATOMS and QUARKS surveys, respectively.
\label{fig_targets} }
\end{figure*}

Observations show 
that high-mass stars are rare, 
and mostly born in  massive clumps within the Galactic plane 
that are much denser than low-mass star forming clouds \citep[e.g.,][]{2009A&A...504..415S}. Considering their large distances ($\sim$kpc) and high dust extinction, studies of those massive clumps need high resolution interferometric observations that can resolve their internal gas structures and kinematics.  The Atacama Large Millimeter/submillimeter Array (ALMA) 
provides an opportunity to investigate the inner hierarchical structures 
of high-mass star forming regions in great detail.
There is growing evidence of the existence of filaments within massive clumps 
giving rise to hub-filamentary structures (HFSs)
\citep[e.g.,][]{2013A&A...555A.112P,2018ApJ...852...12Y,2022A&A...658A.114K,2023MNRAS.522.3719L,2023ApJ...953...40Y,2023ApJ...950..148M}.
However,  most of these have been the results of case studies.
The `ALMA Three-millimeter Observations of Massive Star-forming regions' (ATOMS)   survey \citep{2020MNRAS.496.2790L} 
observed 146 active star-forming regions with ALMA at Band 3 ($\lambda\sim$ 3 mm),
aiming to systematically investigate the gas distribution, stellar feedback, and filaments inside massive clumps.
The ATOMS survey revealed, in a statistical way, that HFSs
are common within massive clumps  \citep{2022MNRAS.514.6038Z}. 
Case studies of HFSs  based on the ATOMS survey \citep{2022MNRAS.510.5009L,2022MNRAS.511.4480L,2023MNRAS.520.3259X}
confirmed that HFSs
at scales from 0.1 pc to several pc play a key role in high-mass star formation, and the stellar feedback
would in turn regulate the  shapes and  evolution of HFSs 
\citep[e.g.,][]{2013A&A...555A.112P,2022A&A...658A.114K,2023MNRAS.522.3719L}.
A higher-resolution ($\sim$1000 au) survey would enable to systematically establish whether the 
hierarchical structures remain common down to a protocluster scale, and to provide for theoretical studies the observational constraints onto the detailed process of gas mass transport down to the protocluster scale.

Inspired by the ATOMS survey, 
we initiated the `Querying Underlying mechanisms of massive star formation with ALMA-Resolved gas
Kinematics and Structures (QUARKS)’ 
survey programme at ALMA (PIs: Lei Zhu, Guido Garay and Tie Liu).
The QUARKS survey selected the densest kernels of the massive clumps in the ATOMS survey, 
forming an unbiased sample (139) of protoclusters, and observed them at Band 6 ($\lambda\sim 1.3$ mm) of ALMA 
with much improved angular resolution than the ATOMS survey.
The QUARKS survey aims (1) to statistically
investigate the star formation process (e.g., fragmentation, outflows, disks) within 
an unbiased sample of protoclusters, (2) more specifically,  
to explore how (hub-)filamentary structures feed individual  protostars within protoclusters, 
(3) to investigate the physical and chemical evolution of dense cores, and (4) to study various feedback mechanisms within protoclusters.

In this paper, we present an overall description of the  QUARKS survey.
The paper is structured as follows: In Section \ref{secsampleandobs}, we introduce the source sample and ALMA observations
of the QUARKS survey. In Section \ref{sec_dr}, we describe the data reduction processes of both the continuum maps and spectral cubes. 
In Section \ref{sec_sgrb2},  a first look at Sgr B2(M), a source within the sample of this survey, is presented
to  check the data quality and to  explore in a preliminary fashion how the data can be linked to 
some of the science objectives of the QUARKS survey. 
In Section \ref{sec_stopic}, we compare the QUARKS surveys with other ALMA programs and present
more science topics that can be addressed by the QUARKS survey. A summary is provided in Section \ref{secsummary}.


%

\section{Sample and observations} \label{secsampleandobs}

\subsection{Sample}\label{sec_sample}

The QUARKS sample contains 139 protoclusters, selected 
as the densest kernels of high-mass star forming regions
revealed by the 
ATOMS survey \citep{2020MNRAS.496.2790L}.
Among the 146 sources of the ATOMS, 
two sources are low-mass clumps ($<15$ $M_\odot$), four sources are dominanted by
extended H{\sc ii} regions with angular sizes larger than the primary beam at Band 6, 
and one source has no detection of continuum emission by  ATOMS. 
The seven sources show no massive and compact kernels under the view of ATOMS and 
were thus excluded in the QUARKS sample.
The ATOMS sources were
selected as the bright sources ($T_{\rm b}>2$ K) of the CS $J=2-1$ survey of  \citet{1996A&AS..115...81B}, 
a complete and homogeneous molecular line survey of UC H{\sc ii} region candidates in the Galactic plane.
The ATOMS sample 
was also observed in the SIMBA survey, a 1.2 mm continuum emission survey using the SEST telescope by \citet{2004A&A...426...97F},
and the survey of HCN (4-3) and CS (7-6) using the ASTE telescope by \citet{2016ApJ...829...59L}.
The ALMA primary beam size at Band 6 is about 0.4 times the value in Band 3, hence 
the filed of view (FoV) of the QUARKS survey can only cover part of the FoV of the ATOMS survey (Figure \ref{fig_targets}).
For 17 sources with elongated and filamentary structures revealed by the 3 mm continuum
emission of the ATOMS survey, 
two pointings for each were requested to entirely cover their dense kernels (e.g., 
IRAS 08448-4343 as shown in
the right panel of Figure \ref{fig_targets}). 
In total, QUARKS has 156 single-pointings to observe 139 protoclusters.
The clump masses ($M_{\rm clump}$), quoted from \citet{2020MNRAS.496.2790L},  range from 40 to 2.5$\times$10$^{5}$ $M_{\sun}$.
These sources are located on the Galactic plane within a range of $|b|<2.1\degr$ and 
$-98.4\degr <l<43.8\degr$, providing an unbiased sample of protoclusters in the   inner Galactic plane.
For all the targets of the QUARKS survey, the target names and
target centers (in both Equatorial and Galactic coordinate systems) are listed in Table \ref{tab_targets}.

\begin{table}[!htb]
\centering
\caption{ALMA configurations of the QUARKS survey. \label{tabobsconfig}}
\begin{tabular}{lccc}
\hline\hline
Configuration$^{(1)}$  &    Resolution & MRS &  rms$_{\rm line}^{(2)}$  \\
              &         (arcsec)      &          (arcsec)  &    (mJy beam$^{-1}$) \\
\hline
ACA           &  4.5 &27  & 40 \\
12-m array C-2           &   1& 9&6\\
12-m array C-5           &  0.25   & 3.5&4\\
combined           &     0.3 & 27  & 3\\                 
 \hline
\end{tabular}{\flushleft
$^{(1)}$ For a small fraction of the sources, the moderate-resolution observations were conducted by the C-3 (instead of  the C-2) configuration,
and the high-resolution  observations were conducted by the C-6 (instead of  the C-5) configuration.\\
$^{(2)}$ The precise values of the line sensitivities (rms$_{\rm line}$) 
of different sources vary  by $\sim$50 percent because of factors such as the
integration time  and calibration.
\\
}
\end{table}

\begin{table*}[!thb]
\centering
\caption{Set up of the spectral windows and the main targeted lines in the QUARKS survey.\label{tab_spws}}
\begin{tabular}{cccccccl}
\hline
\hline
SPW & $f_{\rm center}$ &  Bandwidth$^{(1)}$ & $\delta V$ & Species & Trans.$^{(2)}$& $E_{\rm u}$ & Note\\
    & (GHz)             &  (GHz)     & km s$^{-1}$&             &          & (K)   &  \\
\hline
SPW1& 217.918429        & 1.875      & 1.344      &  SiO          &    5-4   & 31.2 & Jet/shock/outflow tracer  \\
    &                   &            &            & H$_2$CO       &    3-2    &  20--68 & Temperature/infall/outflow tracer         \\
    &                   &            &            & HC$_3$N       &    24-23  & 131  & Infall/dense-gas/filament tracer        \\
    &                   &            &            & CH$_3$OH      &    4$_2$-3$_1$& 45.5    & Tracer of temperature/COMs        \\
SPW2& 220.318632        & 1.875      & 1.329      &   $^{13}$CO/C$^{18}$O   &    2-1&     16  & Filament/outflow tracer\\
    &                   &            &            & CH$_3$CN      &    12-11 &  $>$68  & Hot-core/massive-disk tracer        \\
    &                   &            &            & SO            &    6-5 &   35 & High-density/shock/outflow tracer        \\
SPW3& 231.369566        & 1.875      & 1.266      &    CO         &    2-1 &  16.6    & Ambient-gas/outflow tracer \\
    &                   &            &            & N$_2$D$^+$    &    3-2  &  22.2   & Cold-gas tracer        \\
    &                   &            &            & $^{13}$CS    &    5--4  &  33.3   & Ambient gas/filament tracer\\
    &                   &            &            & H$^{+\ (3)}$   &H$_{30}\alpha$&  &Ionized-gas/HII-region tracer\\
SPW4& 233.519748        & 1.875      & 1.254      & NH$_2$CHO      &  11-10  &   $>$94        & Hot-core tracer \\
    &                   &            &            & C$_2$H$_5$CN    &    26-25     &   $>$100  &  Hot-core tracer  \\
    &                   &            &            & SO$_2$    &    --$^{(5)}$ &   $>$200  & Tracer of sulfur-rich hot gas  \\    
    &                   &            &            & Continuum$^{(4)}$    &                &   & Dust tracer        \\
\hline
\end{tabular}{\flushleft
$^{(1)}$ The bandwidths listed here are for the 12-m-array observation. The bandwidth of each SPW of the ACA observations is 2 GHz.\\
$^{(2)}$ The transitions of C$_2$H$_5$CN, NH$_2$CHO, and CH$_3$CN referred to line groups. The H$_2$CO has more than one transtions
in SPW1.  For these species, the ranges of $E_{\rm u}$ are listed. \\
$^{(3)}$ H${_n\alpha}$ is the transition of H atom but traces H$^+$ \citep[e.g.,][]{2002ASSL..282.....G,2023AA...671L...1L}. \\
$^{(4)}$ The SPW4 lacks low-energy lines and thus is good for extrating continuum emission (Section \ref{sec_continuumreduct}) and tracing 
hot cores (see Section \ref{sechotcores}).\\
$^{(5)}$ Transitions of SO$_2$ such as $16_{6,10}-17_{5,13}$ and $28_{3,25}-28_{2,26}$ are covered by the SPW4.\\
}
\end{table*}

\begin{figure}[!hbt]
\centering
\includegraphics[width=0.991\linewidth]{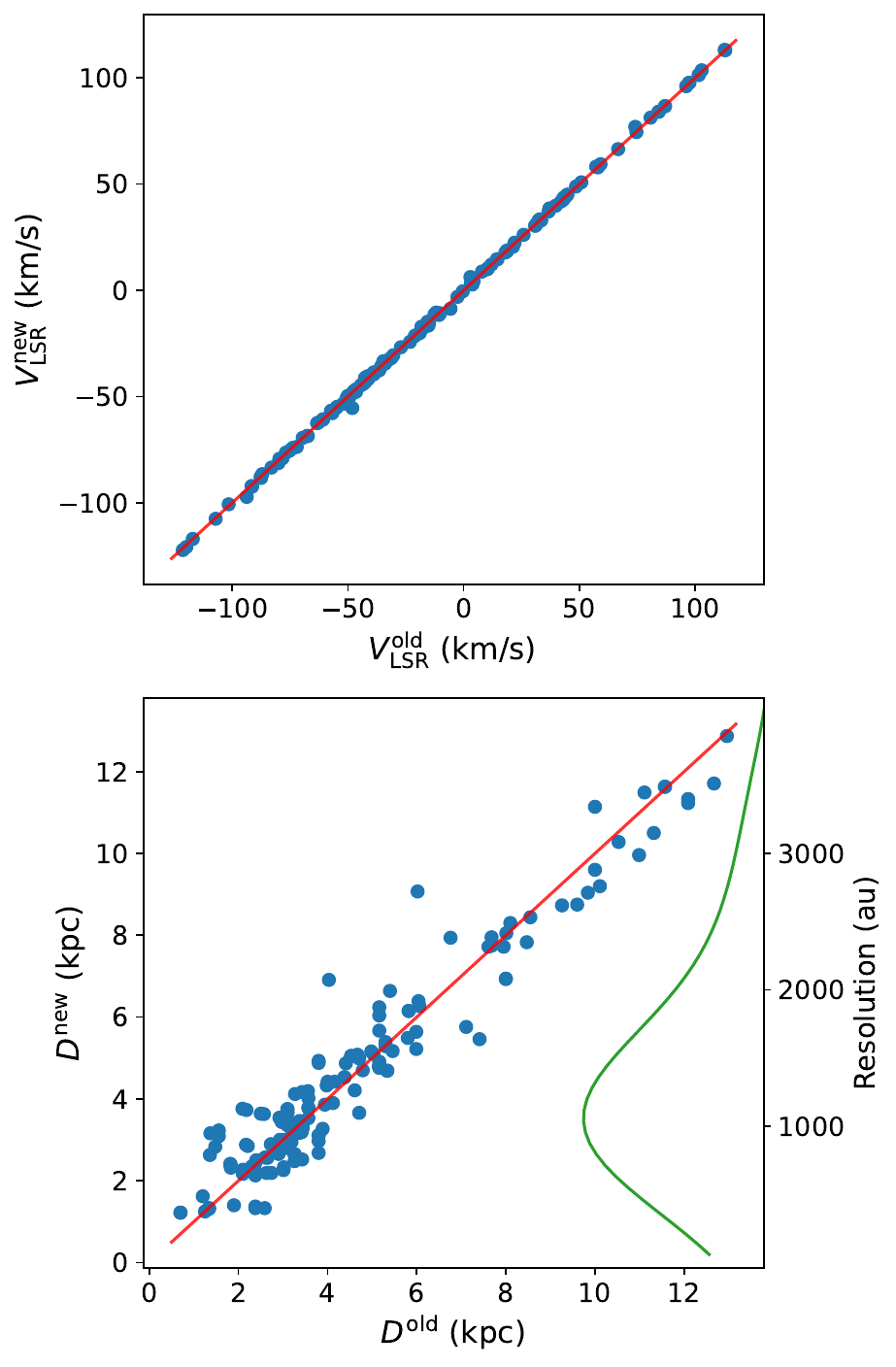}
\caption{Comparison between $V_{\rm LSR}^{\rm new}$ and $V_{\rm LSR}^{\rm old}$ 
is shown in the upper panel and
between $D^{\rm new}$ and $D^{\rm old}$ (Lower) in the lower panel. The red lines represent
$y=x$. The green curve represents the kernel distribution estimation 
of the source distances (with respect to the left Y axis) or of the Band-6 spatial resolution 
(with respect to the right Y axis) of the QUARKS sample.   \label{fig_dist}
}
\end{figure}

\subsection{Re-evaluation of sample distance}
The distances of the QUARKS sample ($D^{\rm old}$) adopted by the ATOMS survey were
initially derived from the
velocities of CS $J=2-1$ with an old Galactic rotation curve model \citep{1996A&AS..115...81B}.
We updated the distances ($D^{\rm new}$) using the distance calculator of \citet{2016ApJ...823...77R},
based on the new systemic velocities of H$^{13}$CO$^+$ ($V_{\rm LSR}^{\rm new}$) measured from the 
ATOMS survey \citep{2020MNRAS.496.2821L}.
The distance calculator constrains the probability density function (PDF) 
of distance based on four types of information: 
kinematic distance (KD), the spiral arm model (SA), Galactic latitude (GL), and parallax source (PS) with different 
weights, which were assumed to be
0.85, 0.15, 0.85, 0.15, respectively \citep{2016ApJ...823...77R}.
The velocities  and
distances  of the sample are listed in Table \ref{tab_targets}.

As shown in Figure \ref{fig_dist}, the $V_{\rm LSR}^{\rm old}$ and $V_{\rm LSR}^{\rm new}$ are 
consistent with each other, and the 
standard deviation of their differences is $\sim$1 km s$^{-1}$, much smaller than the
virial velocity ($\sigma_{vir}$) of $\sim$ 5 km s$^{-1}$ of giant molecular clouds \citep{2023ASPC..534....1C}. 
Thus, the new calculated distance should not be influenced much by the new velocity employed.
Moreover, we compared $D^{\rm new}$ and $D^{\rm old}$, and found that
they are consistent with each other with an 1-$\sigma$ deviation of 0.7 kpc
(Figure \ref{fig_dist}).
The distribution of the QUARKS survey linear resolutions (estimated adopting an angular resolution of 0.3\arcsec; 
see Section \ref{sec_obs}) 
is shown as the 
green line in the lower panel of Figure \ref{fig_dist}. 
The distances ($D^{\rm new}$) of the sample of the QUARKS survey range from 1.2 to 12.9 kpc.
The median value of distances is 3.7 kpc, at which a linear resolution 
of 1100 au can be achieved. 
The resolutions for most sources are within the range of 500--2000 au (Figure \ref{fig_dist}).

\subsection{ALMA Observations} \label{sec_obs}
We conducted single-pointing observations for the 156 targets of the QUARKS survey using the ALMA.
The observations started from late October 2021
(Project IDs: 2021.1.00095.S, 2022.1.00298.S and 2023.1.00425.S; PI: Lei Zhu) 
with both the Atacama Compact 7-m Array (ACA) and the 12-m array of the ALMA
at Band 6.
Observations with the ACA were completed in late May 2022.
Observations with the 12-m Array have been partly executed in Cycles 8 and 9 and 
are expected to be completed in ALMA Cycle 10 (from October 2023 to September 2024).

\begin{figure*}[!thb]
\centering
\includegraphics[width=0.99\linewidth]{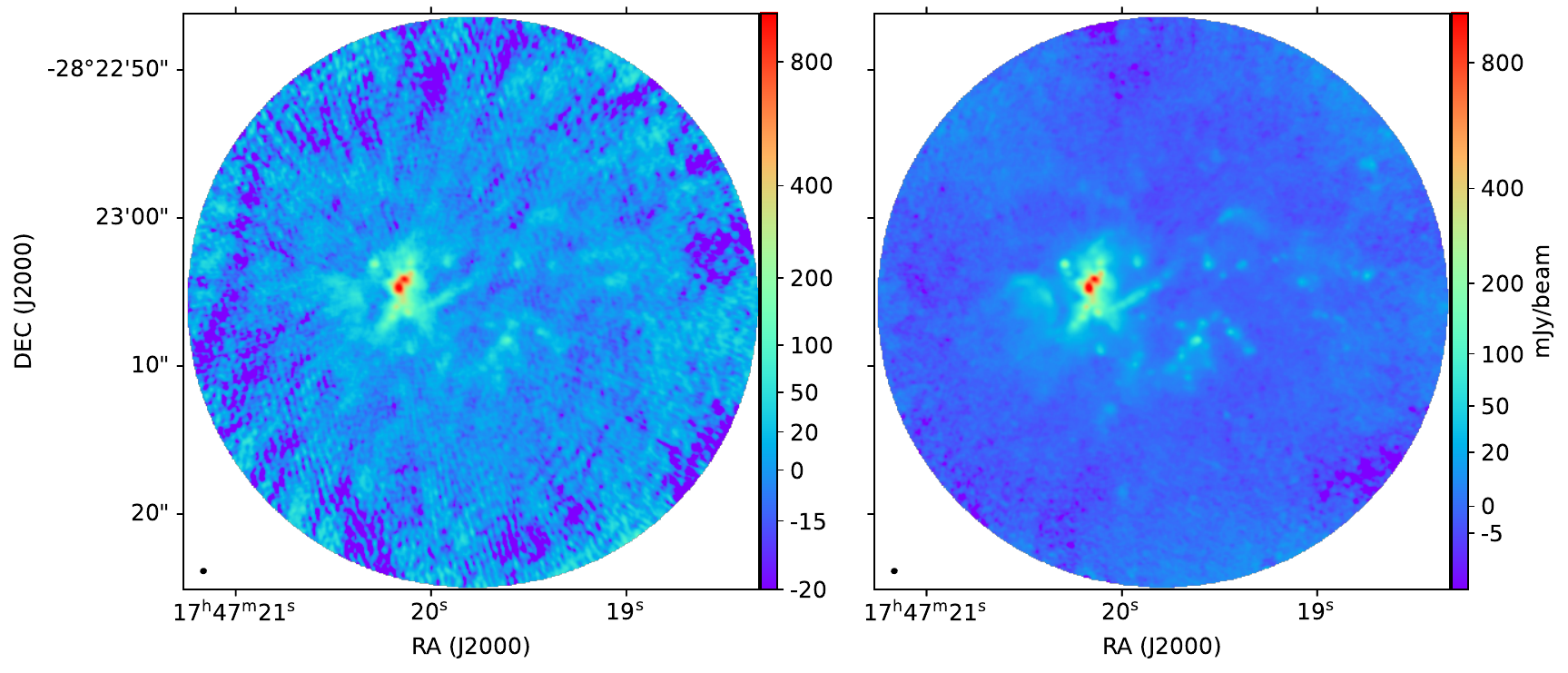}\\
\includegraphics[width=0.94\linewidth]{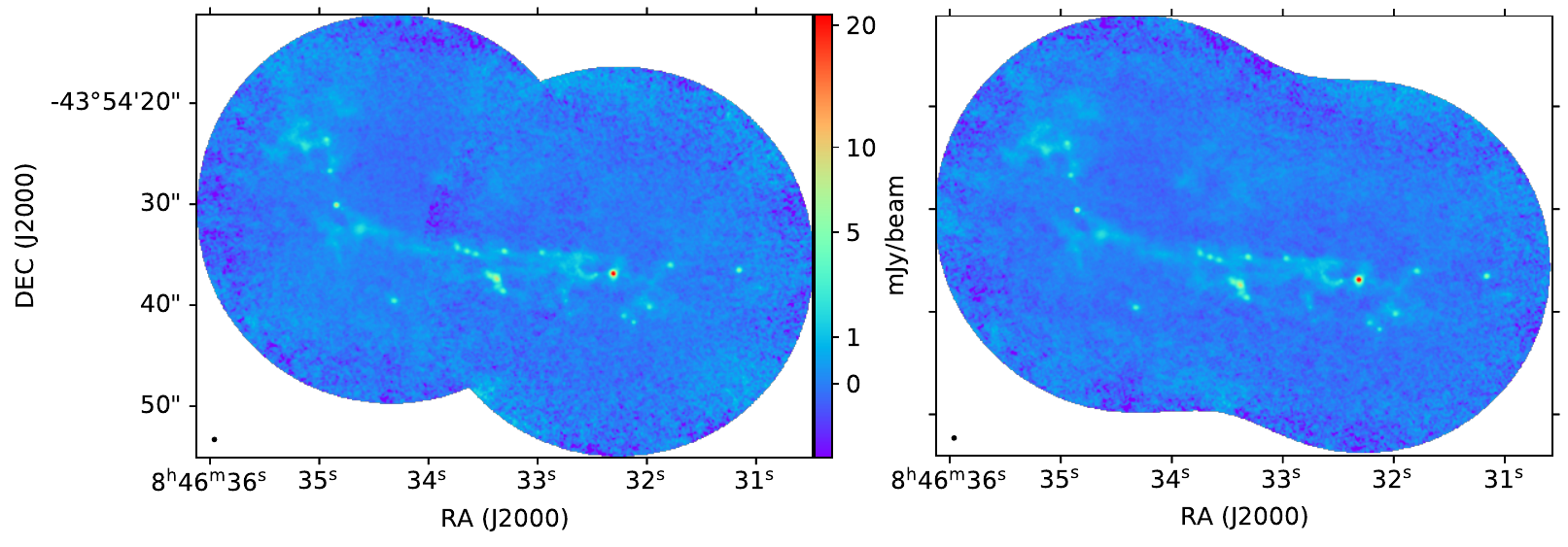}\hspace{0.055\linewidth}
\caption{The ALMA Band 6 continuum of Sgr B2(M) (upper) and IRAS 08448-4343 (lower).
For Sgr B2(M), the upper right and upper left panels are the images 
with and without selfcalibration, respectively.
For IRAS 08448-4343,  the lower left panel shows the  image stitched from two independently cleaned images,
and the lower right panel shows the  the image cleaned from the combined visibility data.
The two lower panels share the same colorbar.
In each panel, primary beam correction has been applied, and
the black small ellipse in the lower-left corner represents the synthesized
beam. \label{fig_sgrb2_compareselfcal}}
\end{figure*}

\begin{figure*}[!thb]
\centering
\includegraphics[width=0.999\linewidth]{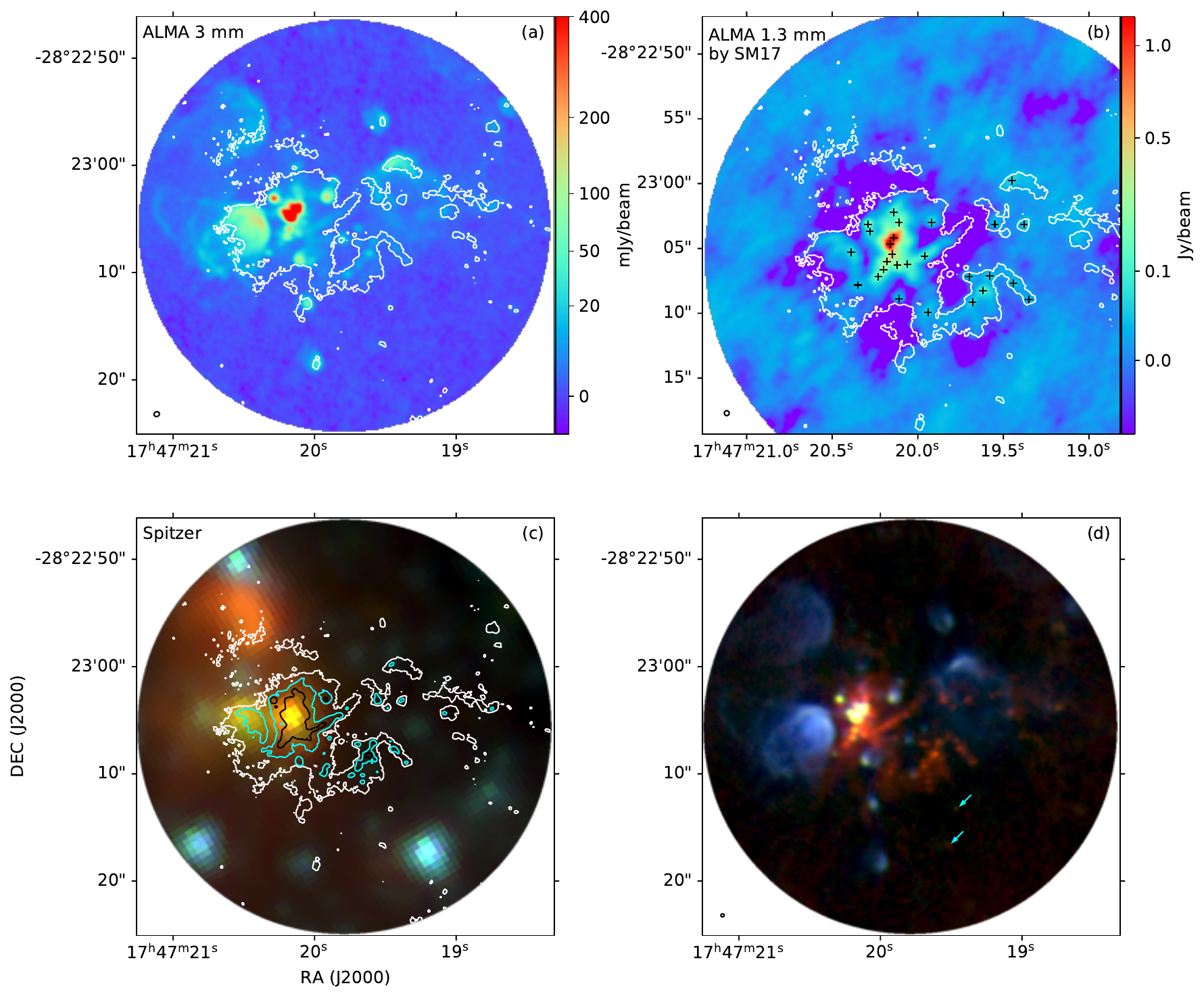}
\caption{(a) Image of the ALMA Band 3  continuum emission of Sgr B2(M) with a resolution of $\sim 0.5\arcsec$ by \citet{2018ApJ...853..171G}. 
The beam size  is shown as an ellipse at the lower-left corner. The white contour shows the
ALMA Band 6 (at 1.3 mm) continuum emission from the QUARKS survey at the level of  5$\sigma$ (2.5 mJy beam$^{-1}$). 
(b) Image of the ALMA Band 6 continuum emission  from SM17 \citep{2017A&A...604A...6S}.
The fields of view of SM17 and the QUARKS survey are not identical,
but all 27 cores identified by \citet{2017A&A...604A...6S} marked by the black crosses 
are shown in this panel.
Contour is the same as in panel (a).
Note that the color map ({\it rainbow}  with a power-law normalization with a power index of 0.4) 
is similar to that of 
Figure \ref{fig_sgrb2_compareselfcal}, where the  ALMA Band 6 continuum  of the QUARKS survey is shown.
(c) Three-color image composed by the Spitzer 8 (red), 4.5 (green), 3.4 (blue) $\mu$m continuum. 
The white, cyan and black contours show the ALMA Band 6 continuum of the QUARKS survey, at the levels
2.5, 15, and 50 mJy beam$^{-1}$, respectively. (d) Three-color image composed of the
ALMA Band 6 continuum of the QUARKS survey in red, the ALMA Band 3 continuum  by \citet{2018ApJ...853..171G} in green,
and the VLA Band C (at $\sim$6 GHz) continuum by \citet{2022A&A...666A..31M} in blue. 
The cyan arrows pinpoint Cores 55 and 96 (see Figure \ref{fig_sgrb2cores} and Section \ref{sec_selfcali_check}).
The beam size of the QUARKS survey is shown as an ellipse at the lower-left corner.
\label{fig_colorimages} }
\end{figure*} 

\begin{figure*}[!thb]
\centering
\includegraphics[width=0.9\linewidth]{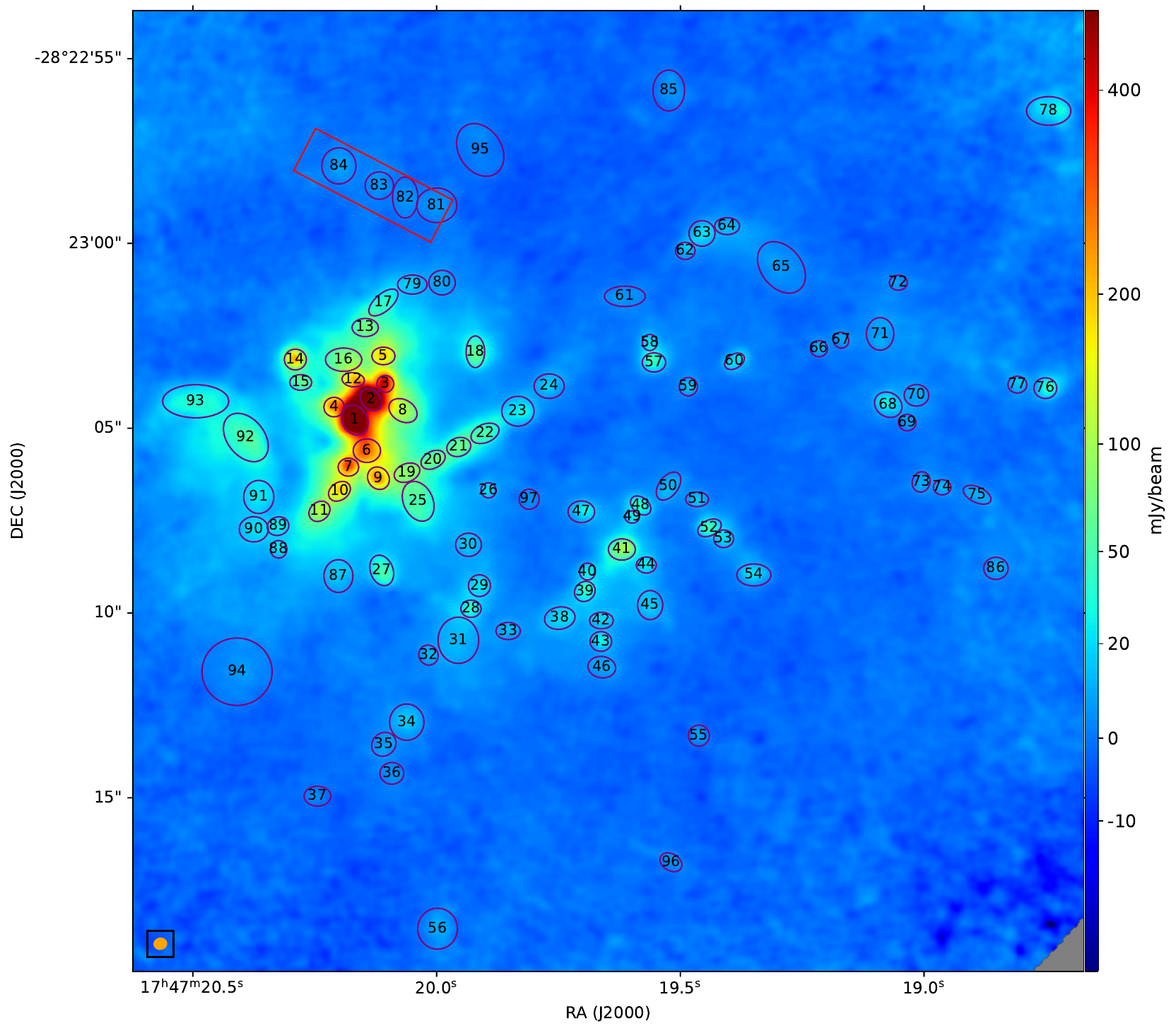}
\caption{Zoom in image of the right panel of Figure \ref{fig_sgrb2_compareselfcal}.
The purple ellipses and black numbers marked the cores and substructures of Sgr B2(M) detected in the Band-6 continuum image
(Section \ref{sec_srgb2cores}).
The red rectangle marks the north arc of the Sgr B2(M).
The orange  ellipse in the lower-left corner represents the synthesized
beam size. \label{fig_sgrb2cores}}
\end{figure*}

\begin{figure}[!thb]
\includegraphics[width=0.9\linewidth]{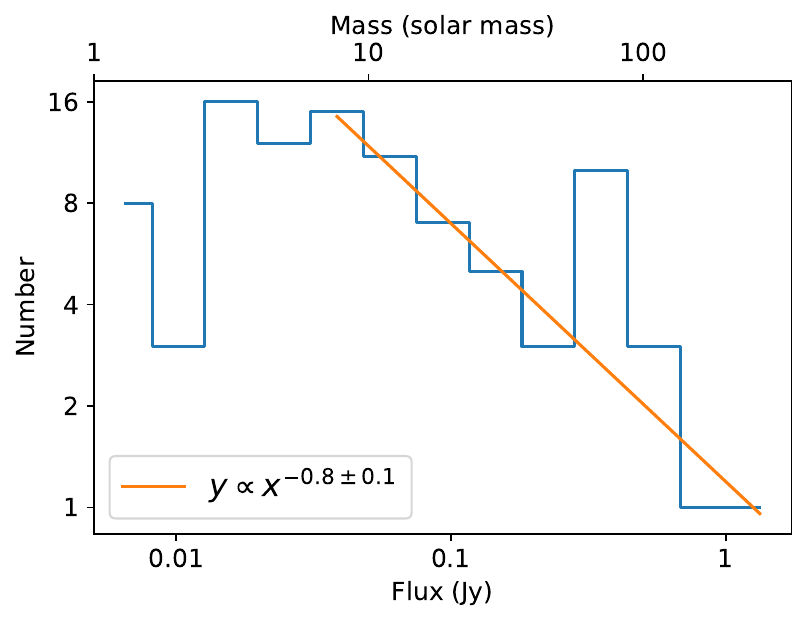}
\includegraphics[width=0.92\linewidth]{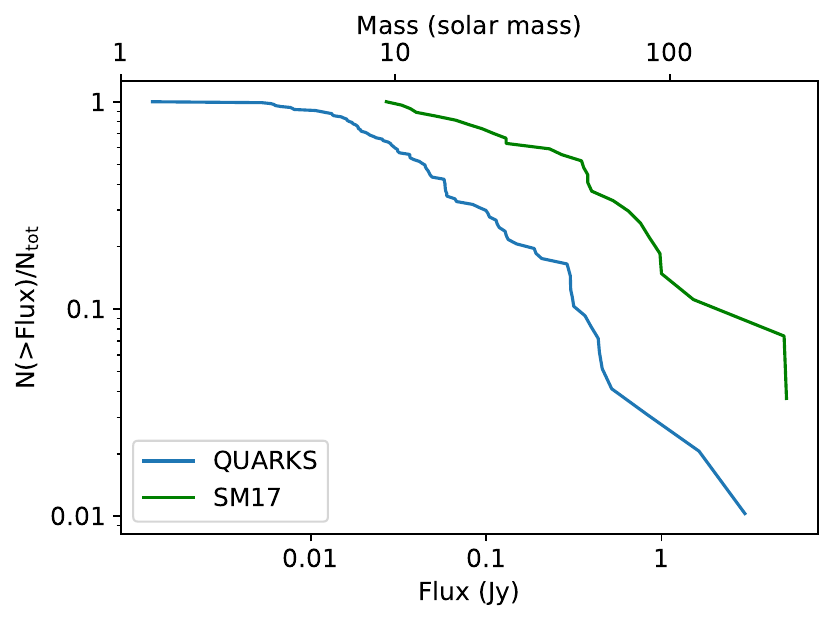}
\includegraphics[width=0.92\linewidth]{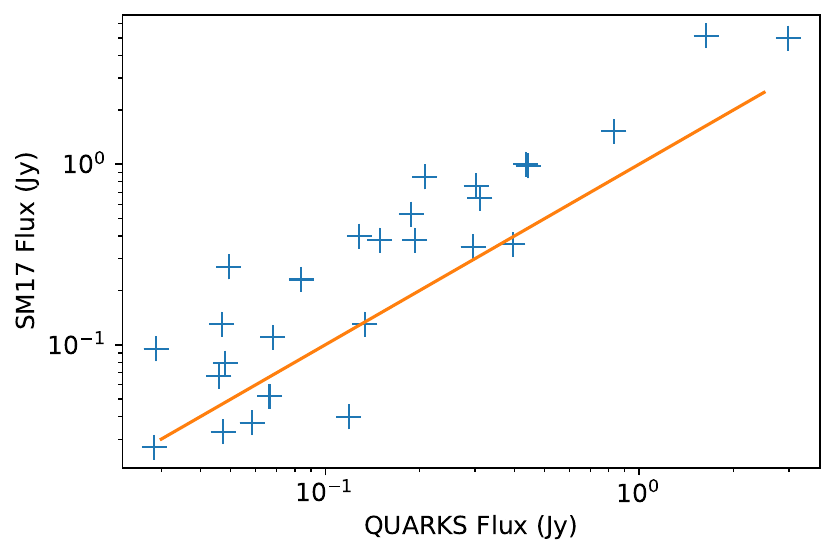}
\caption{Upper: The fluxes distribution of the cores in Sgr B2(M) (Figure \ref{fig_sgrb2cores}).
The orange line represents the power-law fitting of data at the high-flux end.
Middle: The cumulative  fluxes distribution of the cores in Sgr B2(M) by QUARKS (blue) and SM17 \citep[green;][]{2017A&A...604A...6S}.
For reference, the corresponding mass estimate as a first order of approximation (Section \ref{sec_srgb2cores}) is displayed at the top axis of the
upper and middle panels.
Lower: Comparison between the core fluxes measured by QUARKS and SM17. 
The orange line represents $y=x$.
Note that QUARKS detected more fragmented cores than SM17 because of better angular resolution and sensitivity (Section \ref{sec_srgb2cores}). 
\label{fig_cmf}}
\end{figure}

For each source, observations in three different ALMA configurations were proposed.
The observations in low, moderate and high resolution configurations   were conducted by the ACA,
ALMA 12-m array C-2 and C-5 configurations, respectively.
In the ACA, C-2 and C-5 observations, the on-source time for one science target
was approximately 5, 1, and 5 minutes, respectively. 
The typical values of the  
angular resolution, maximum recovering scale (MRS), and the sensitivity
of spectral lines (rms$_{\rm line}$) for different configurations, 
including the combined configuration, are summarized in Table \ref{tabobsconfig}. 
The continuum sensitivity (rms$_{\rm cont}$) is not listed,
because the  rms$_{\rm cont}$ of the combined data varies in a wide range from 0.1 to several mJy
beam$^{-1}$ for different sources,
mostly because of the phase calibration and dynamic range limitations. 
Self-calibration was conducted for sources containing strong continuum (with peak value $\gtrsim 50$ mJy beam$^{-1}$) cores (Section \ref{sec_dr}).
We note that for 1.3 mm continuum 1 mJy beam$^{-1}$ corresponds to 
a gas mass of 0.3 $M_{\sun}$ at a distance of 5 kpc, 
derived assuming a median dust temperature of 30 K, a dust opacity per unit dust mass of 1 cm$^{2}$ g$^{-1}$ 
at 1.3 mm \citep{1994A&A...291..943O}, and a gas-to-dust mass ratio of 100. 
The angular resolutions of the combined images are typically better than 0.35 arcsec,  
which is about one-sixth of the resolution  of the
ATOMS survey (typically better than 2 arcsec) in Band 3 \citep{2020MNRAS.496.2790L}. The significantly improved angular resolution of the QUARKS survey 
enables us to resolve dense cores, including massive starless core candidates and protostellar ones
(star-forming dense cores, hot molecular cores and UC H{\sc ii} regions) within protoclusters,  
down to a physical scale of 3500 au in the most distant sources at a distance of $\sim$10 kpc.
For sources at a distance smaller than 3 kpc, the QUARKS survey 
can achieve a linear resolution better than 1000 au, 
and thus enable us to search for massive disk candidates \citep[e.g.,][]{2015ApJ...813L..19J,2016ApJ...823..125C,2022ApJ...929...68O}. 

The observations employed the Band 6 receivers in dual-polarization mode.
Four spectral windows (SPW 1--4) were configured, and
the frequency setups are shown in Table \ref{tab_spws}.
The second, third and fourth columns of Table \ref{tab_spws} list the central frequency, bandwidth and velocity resolution ($\delta V$)
of each SPW (column 1). For the observations of the 12-m array and the ACA, each of the four SPWs has
a bandwidth of 1.875 GHz and 2 GHz, respectively. 
The 5th to 7th columns of Table \ref{tab_spws} list the parameters of the main transitions covered by
each SPW, including the species name, transition labels, and upper-level energy ($E_{\rm u}$).
The four SPWs were configured to cover some commonly used lines 
including tracers for ambient gas (e.g., CO, $^{13}$CO, C$^{18}$O), cold gas (N$_2$D$^+$),
filament and infall  (e.g., HC$_3$N), hot core (e.g., CH$_3$OH, C$_2$H$_5$CN,
NH$_2$CHO),
ionized gas  (H$_{30}\alpha$), shocked gas (e.g., SiO, SO, H$_2$CO),
and massive protostellar disk (CH$_3$CN), as listed in Table \ref{tab_spws}.
SPW4 does not cover any low-energy  strong lines (with upper-level energies $E_{u}<100$ K) 
and thus is good for measuring the continuum emission.
The Band 6 continuum is mainly produced by the dust emission.
However, the Band-3 continuum emission of the ATOMS survey 
arises from a blend of thermal dust emission and free-free emission from ionized gas \citep{2023MNRAS.520.3245Z}.
Combining the data of the QUARKS and ATOMS surveys, the dust emission and the free-free emission
can be partly decoupled.

\begin{figure*}[!thb]
\centering
\includegraphics[width=0.99\linewidth]{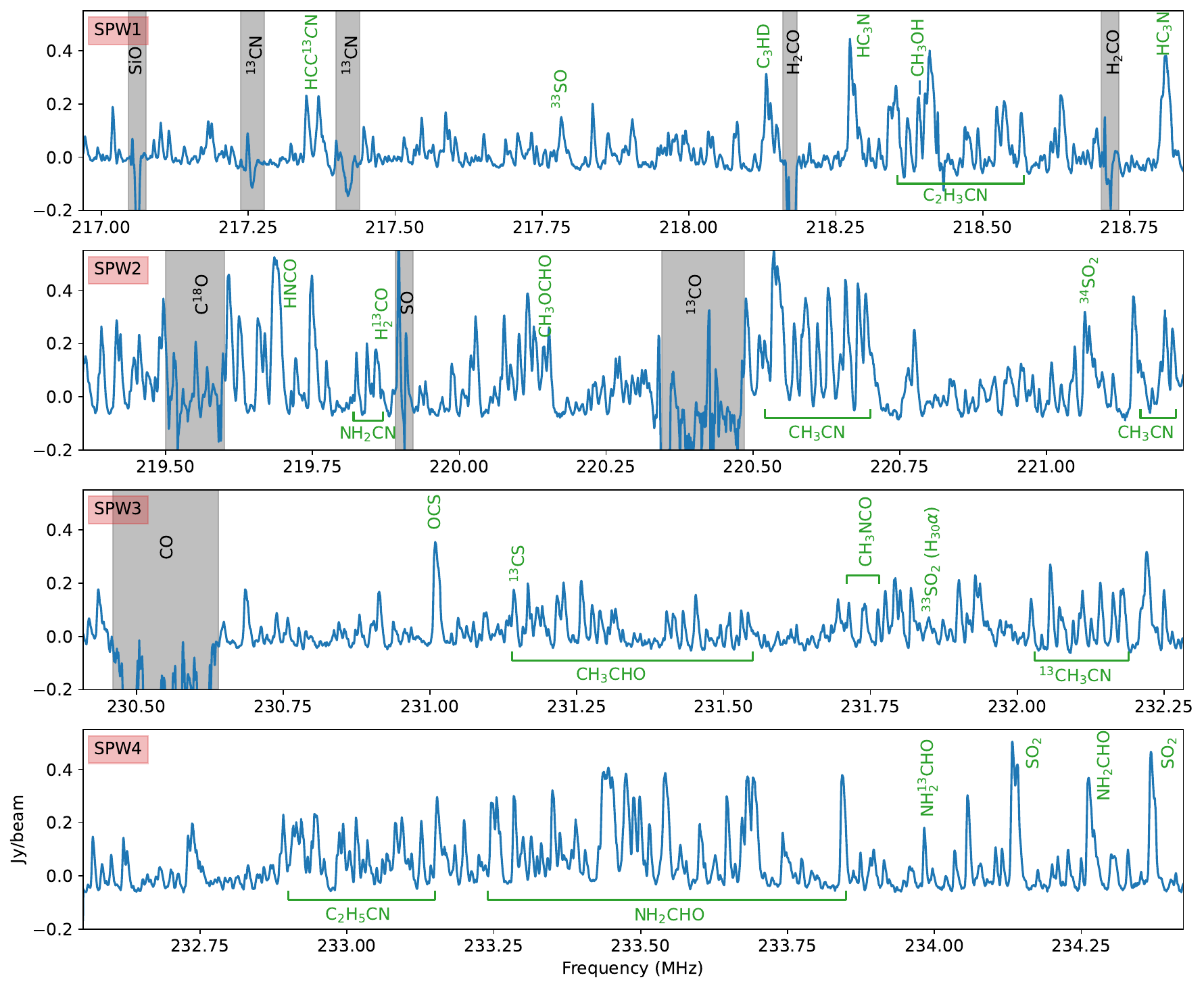}
\caption{The QUARKS spectra of core 6, the line-richest core among those in Figure \ref{fig_sgrb2cores}.
The gray shaded regions mark the strong lines showing absorption. The species
of strong emission line or group of lines are labeled in green. 
Each of the labeled line groups is assigned to one species that dominates the spectral emission within the corrsponding frequency range. 
The velocity ($V_{\rm LSR}$) of core 6 is 67 km s$^{-1}$. 
 \label{fig_core6spe}}
\end{figure*}

The channel widths of the four SPWs are 976.563 kHz, corresponding to a $\delta V$
of 1.25--1.35 km s$^{-1}$, depending on the rest frequency.
The spectral resolution is better  than that  of the two wide-band SPWs (SPWs 7 and 8) 
of the ATOMS survey, $\sim$1.6 km s$^{-1}$ \citep{2020MNRAS.496.2790L}. 
The velocity resolution of the QUARKS survey is high enough to spectrally resolve 
the emission lines of the outflows and ionized gas whose  
line wings or line widths are typically larger than 10 km s$^{-1}$
\citep{2021MNRAS.505.2801L,2022MNRAS.511.3618L,2022MNRAS.510.4998Z},
and the emission of complex organic molecules (COMs) from 
hot cores \citep[$\Delta V\sim$ 5 km s$^{-1}$;][]{2022MNRAS.511.3463Q,2023ApJ...950...57T}. 
The velocity gradients along the longest branchs of HFSs 
were systematically revealed in the ATOMS survey to be
8.7 km s$^{-1}$ pc$^{-1}$ on average \citep{2022MNRAS.514.6038Z},
corresponding to a velocity difference of $\sim$5 km s$^{-1}$ at the 
maximum recovering scale (0.6 pc at a distance of 5 kpc) of the QUARKS survey.
Thus,  spectral lines of species such as HC$_3$N can be used to trace the
multi-scale structures and gas kinematics within the clumps \citep{2022MNRAS.511.4480L,2023MNRAS.520.3259X}.
The massive starless core candidates found so far tend to show no emission lines or 
narrow lines (with linewidths $<1$ km s$^{-1}$) of cold gas tracers at Band 6 \citep[e.g.,][]{2014ApJ...796L...2C,2017ApJ...834..193K}.
The N$_2$D$^+$ data of the QUARKS survey do not have sufficient spectral resolution to fully resolve the 
internal gas kinematics
of possible massive starless cores found by this survey,
but can provide us the spatial distribution of cold gas.
The line widths of the (sub-)Keplerian disks around massive  stars are typically 5--10 km s$^{-1}$ 
\citep{2015ApJ...813L..19J,2016ApJ...823..125C,2019A&A...623A..77S}. 
The velocity resolution of the QUARKS survey  is 
high enough to spectrally resolve the massive disks, 
if detected.
In a word, for most purposes, the spatial and spectral resolution in the QUARKS survey should be high enough for resolving the gas structures and kinematics in massive protoclusters.

\section{Data reduction} \label{sec_dr}
The 12-m-array data and the ACA data of the QUARKS survey were combined to reveal
both compact and extended emission.
Sgr B2(M), also named I17441-2822 in this survey, is one of the QUARKS sources with most complex and richest line emission. It is very complicated to reduce the data and produce high-quality images for this object. In this work, we take Sgr B2(M) as an example for evaluating our data reduction procedure.  
The imaging was carried out using the CASA software package of version 6.5 \citep{2007ASPC..376..127M}. 

\subsection{Flagging of emission line channels}\label{sec_flag}
To separate the emission of continuum and spectral lines, 
the emission line channels need to be flagged. The spectral lines mainly contain the
extended strong emission (e.g., the transitions with $E_{u}<100$ K listed in Table \ref{tab_spws}) 
and the forest of COM lines from hot cores. 
Firstly, we identified all the transitions of strong lines within the 
four SPWs through matching the ALMA pipeline reduced datacubes 
and the laboratory databases for spectral lines \citep[e.g., the CDMS;][]{2001A&A...370L..49M}. 
Since the $V_{\rm LSR}$ of the source is already known (Section \ref{sec_sample}), 
the observed frequencies of the those strong lines can be roughly determined.
Because of the possible existence of multiple velocity components and  of foreground absorption, 
an aggressive strategy was adopted for
strong lines by flagging out
all the channels within $\pm$50 km s$^{-1}$ of them.

For the line forest in  hot cores, we first checked the fits cube of the ALMA pipeline products of the C-5 data,
and manually extracted the spectra of hot cores. We calculated the rms of the unflagged channels, 
and then channels with intensity greater than 5$\times$rms were flagged in our reduction. This procedure is repeated till we can no
longer flag any more channels. Since the influence of the line forest is limited to hot cores,
a relatively conservative strategy was adopted by flagging out the channels of $\pm$10 km s$^{-1}$ around those lines. 
For a source with no fits cube produced by the ALMA pipeline,
the channel flag  of Sgr B2(M) was adopted as a template flag. We shifted the template flag before applying it to other sources according to their velocities 
(Table \ref{tab_targets}).  
If the image quality of a source is not good because of an an improper flag of emission line channels,
we update the flag through checking the hot-core spectra extracted from its fits cube  
obtained in the first round of imaging
and then conducted a second round of imaging.

\subsection{Imaging and self-calibration of continuum data} \label{sec_continuumreduct}
To increase the sensitivity of the continuum image, data of all the four SPWs were used. 
The line-free channels (channels that have not been flagged out by the procedures described in Section \ref{sec_flag}) of both the ACA and 12-m-array data were fed to CASA-{\it tclean}
to make the continuum maps.
The Multi-scale Multi-Frequency Synthesis 
\citep[{\it mtmfs};][]{2007ASPC..376..127M} with an {\it nterm} of 2
 was chosen as the deconvolving algorithm.
This algorithm is appropriate for  cleaning on the combined visibility data of
multiple configurations and multiple SPWs.
The {\it briggs} weighting
with a {\it robust} of 0.5 was adopted.
For a source with strong continuum emission such as Sgr B2(M), self-calibration is necessary to 
improve the image quality (Section \ref{sec_selfcal}).
For a source observed by two pointings such as the IRAS 08448-4343 (see Figure \ref{fig_sgrb2_compareselfcal}), 
we combined the visibility data of the two pointings before imaging.
 
\begin{figure*}[!thb]
\centering
\includegraphics[width=0.49\linewidth]{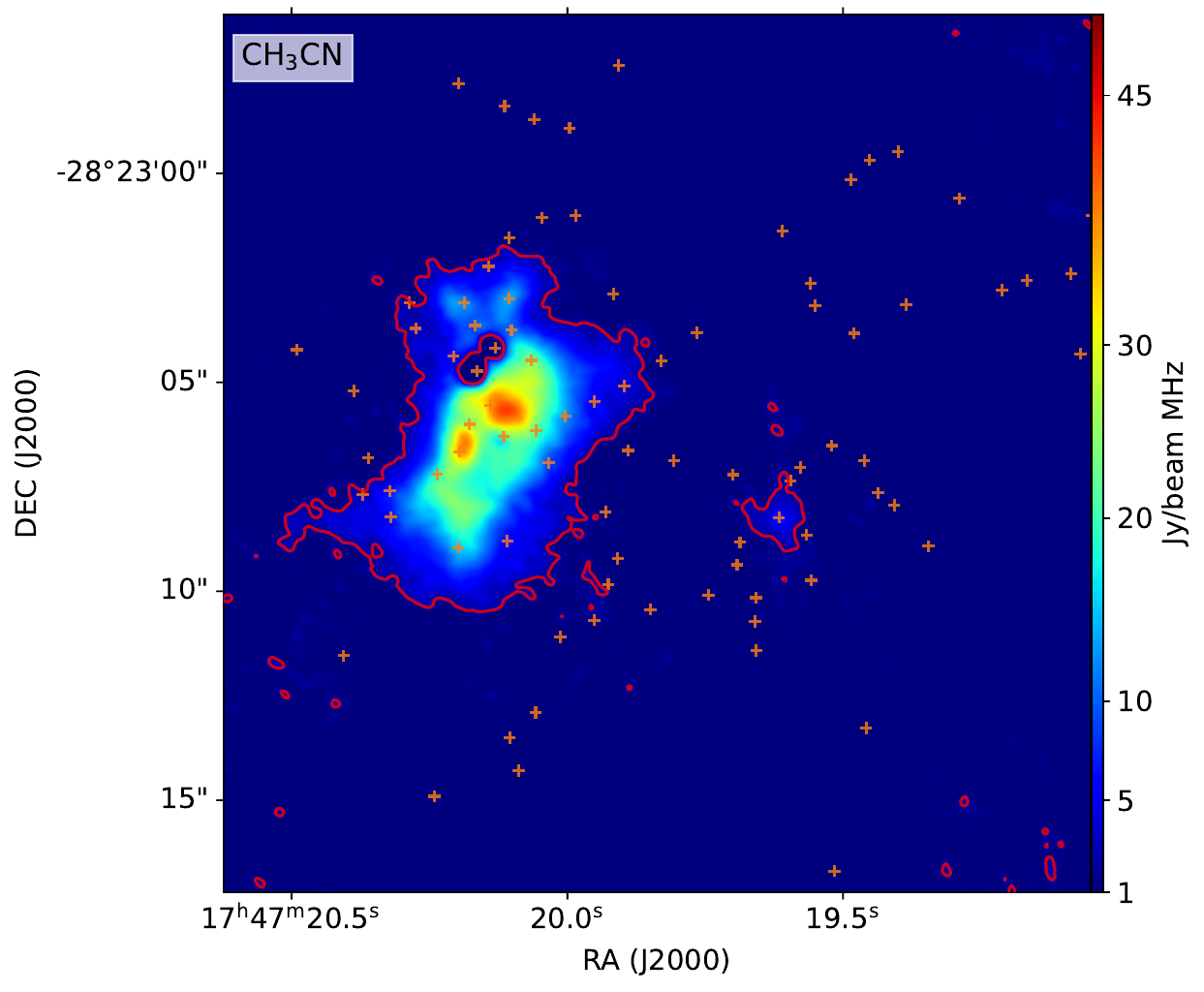}
\includegraphics[width=0.49\linewidth]{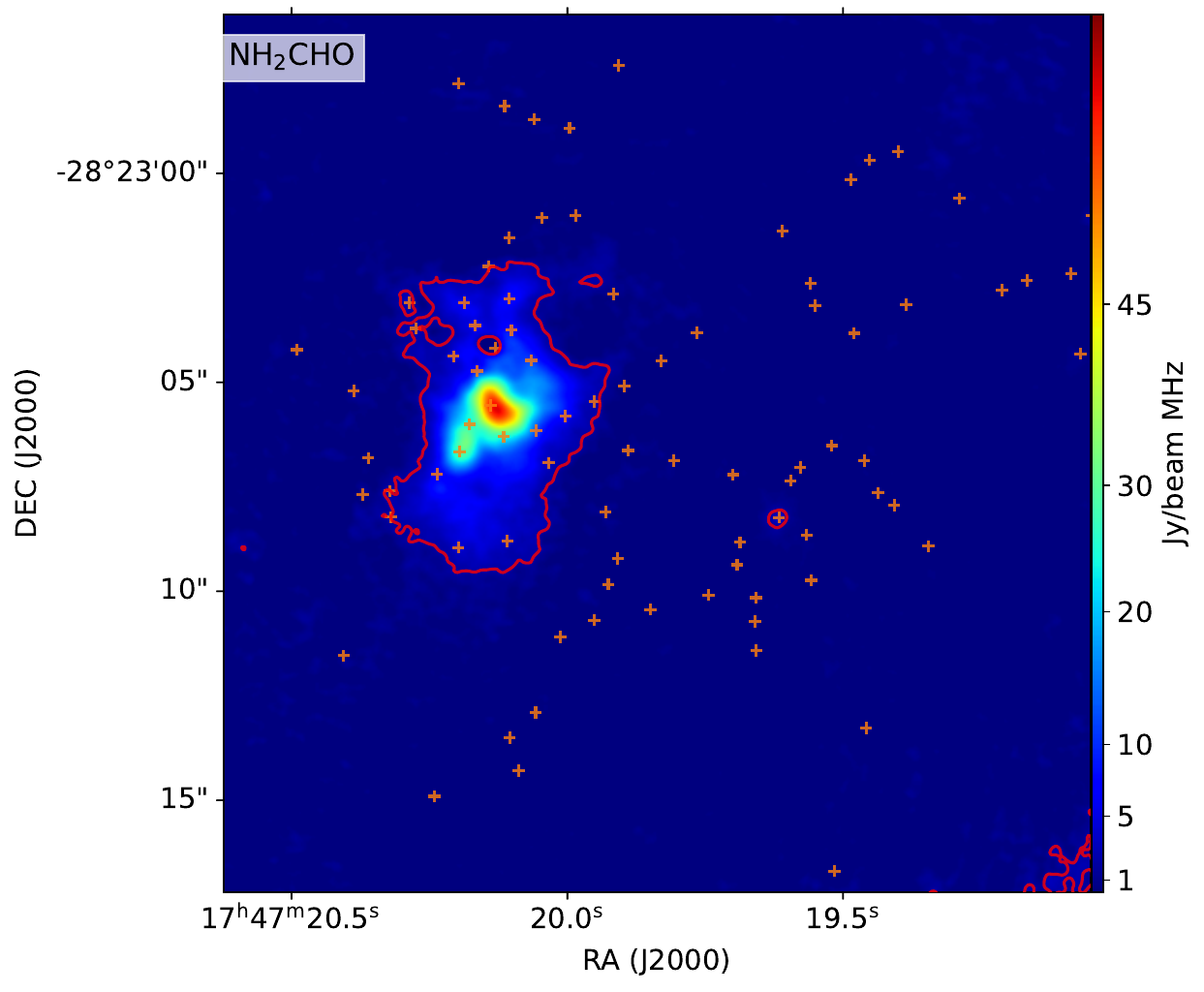}
\caption{Left: The intensity map integrated from 220.52 to 220.7 GHz, which contains
a line group of CH$_3$CN as shown in Figure \ref{fig_core6spe}.
Right:  The intensity map integrated from 233.24 to 233.85 GHz, which contains
a line group of NH$_2$CHO.
In each panel, the red contour is for the intensity map with a level of 2 Jy beam$^{-1}$ MHz,
and the orange crosses mark the continuum cores of this work. \label{CH3CNfig}}
\end{figure*}

\subsubsection{Self-calibration of Sgr B2(M)} \label{sec_selfcal}
The 1.3 mm continuum of the Sgr B2(M) is very strong with a peak intensity ($I_{\rm peak}$) of $\sim$ 1.2 Jy beam$^{-1}$
and a total flux of $>23$ Jy. 
A threshold of 20 mJy beam$^{-1}$ was adopted for the first iteration of the cleaning process 
of the continuum image.  
A lower threshold resulted in divergence. 
With that threshold, the reduced 1.3 mm continuum image of Sgr B2(M) 
show apparent fuzzy artifacts (left panel of Figure \ref{fig_sgrb2_compareselfcal}).
This result agrees with a low dynamic range ($I_{\rm peak}/rms_{\rm cont}\sim 180$) 
for sources having complicated and strong continuum emission, 
implying that the systematic errors of the calibrations of phase and amplitudes dominate over the thermal noise.
Self-calibration is a technique for correcting the visibility phases and/or amplitudes of a source 
by comparing the visibility data with a model of the source itself \citep[e.g.,][]{2022arXiv220705591R}.
For strong continuum sources like Sgr B2(M), self-calibration must be applied 
to improve its dynamic range.

The  phases of the visibility data were self-calibrated based on the model of the first round of cleaning,
following the CASA manual\footnote{\url{https://casaguides.nrao.edu/index.php?title=First_Look_at_Self_Calibration_CASA_6}}. 
A {\it solint} of 20 s was adopted for the {\it gaincal} task.
Adopting a lower threshold (5 mJy beam$^{-1}$ for Sgr B2(M)),
the continuum image was cleaned in the second round, and the model was updated accordingly. 
Then, both  the phases and amplitudes  of the visibility data were self-calibrated based on the updated model.
Finally, adopting an even lower threshold (1.5 mJy beam$^{-1}$ for Sgr B2(M)),
the cleaning process was conducted again to obtain the final continuum image. 
A rms of the residual map of 0.5 mJy was achieved through self-calibration for Sgr B2(M),  corresponding to
a dynamic range of greater than 2000.
The final self-calibrated image greatly improves the quality compared with the non self-calibrated image. 
Extended structures  that can only be  marginally seen in previous observations (Figure \ref{fig_colorimages}) can be clearly recognized
in the self-calibrated continuum image of the QUARKS survey,
e.g, the north arc of Sgr B2(M) 
marked by the red rectangle in Figure \ref{fig_sgrb2cores} 
(see Section \ref{sec_selfcali_check} for the discussion about the reliability of this structure).
Approximately hundred cores or compact substructures 
can be identified in Sgr B2(M) from the Band 6 self-calibrated continuum image of this survey  (see Figure \ref{fig_sgrb2cores} and Section \ref{sec_srgb2cores} for details).
Note that some of compact cores have weak continuum emission, which will be 
carefully scrutinized below for their reliability (see Section \ref{sec_selfcali_check}).

There is no general rule  to choose the thresholds of self-calibration, and
we conducted self-calibration  manually and  iteratively to optimize the thresholds
that can avoid  divergence while yielding good sensitivity. 
The continuum sensitivity of the QUARKS survey 
was expected to be 0.06 mJy beam$^{-1}$, 
according to the proposed observation design (Section \ref{sec_obs}).
The final threshold of the continuum image of Sgr B2(M) is still higher than the 
expected sensitivity, mostly because (1) emission of Sgr B2(M) in Band 6 is dominated by
strong  spectral lines and thus 
the selection of line-free channels is difficult; (2) possible contamination comes 
from  antenna sidelobes;
(3) the strong and complex emission of Sgr B2(M) does not allow 
the self-calibration to further improve the dynamic range.
Here we have conducted only two rounds of self-calibrations and additional rounds with more carefully chosen thresholds may
further improve the image quality. 
For a weak source, such as I08448-4343,  a final threshold of 0.2 mJy beam$^{-1}$  
can be achieved without self-calibration, approaching the proposed $3\sigma$ sensitivity,
and thus self-calibration is not required for it (lower panels of Figure \ref{fig_sgrb2_compareselfcal}).

\subsubsection{Reliability of the self-calibration}\label{sec_selfcali_check}
To evaluate the reliability of our self-calibration,
we examined the self-calibrated continuum image of Sgr B2(M) from the QUARKS survey 
through comparing it with
observations from different frequency bands (Figure \ref{fig_colorimages}).   
Sgr B2(M) was observed with ALMA at Band 6 by SM17 \citep{2017A&A...604A...6S} in Cycle 2, with
a resolution of  $\sim$0.4\arcsec, similar to the resolution  of the QUARKS survey of $\sim0.3$\arcsec.
However, the sensitivity of SM17 is 8 mJy beam$^{-1}$, 
which  is about 16 times lower  than that of the QUARKS survey for Sgr B2(M) ($\sim$0.5 mJy beam$^{-1}$). 
The peak intensities of the Band 6 continuum of Sgr B2(M) are about 1.8 and 1.2 Jy beam$^{-1}$ by 
SM17 and QUARKS, respectively, and are consistent with each other considering the different 
beam sizes and frequencies. 
All the cores of Sgr B2(M) identified by SM17 can be successfully recognized in the QUARKS continuum map 
(Section \ref{sec_srgb2cores}). 

Two isolated cores, numbered 55 and 96 in Figure \ref{fig_sgrb2cores}, 
are visible in  the self-calibrated continuum map of the QUARKS survey,
but would be overwhelmed by noise if self-calibration were not applied (Figure \ref{fig_sgrb2_compareselfcal}). 
These cores are not visible 
in the continuum map of SM17 (see panel (b) of Figure \ref{fig_colorimages}),
but are present  in the ALMA Band 3 ($\lambda\sim3$ mm) continuum map observed by \citet{2018ApJ...853..171G}
with signal-to-noise ratios (S/Ns) greater than 5
at a resolution of $\sim$0.5\arcsec~(see panel (a) of Figure \ref{fig_colorimages}).
The long-baseline  Band-6 observations  by \citet{budaiev2023protostellar} 
have also revealed these two cores at an angular resolution of $\sim 0.06$\arcsec.
Thus, these two cores seen in the self-calibrated continuum map of the QUARKS survey should be reliable. 

Core 97,  with a weak peak intensity of 5 mJy beam$^{-1}$ 
at Band 6 measured from the self-calibrated continuum image of the QUARKS survey, 
is located near the central crowded regions of Sgr B2(M).
This core is not visible on the continuum map of SM17, but 
can be seen on the   ALMA Band 3 \citep{2018ApJ...853..171G}
and VLA Band C \citep[6 GHz;][see also panel (d) of Figure \ref{fig_colorimages}]{2022A&A...666A..31M}
continuum maps. Therefore, core 97 is
considered reliable and could be an embedded UC H{\sc ii} region.

The north arc marked as a red rectangle in Figure \ref{fig_sgrb2cores} has not been reported previously.
We checked the 3 mm continuum \citep{2018ApJ...853..171G} and the 1.3 mm continuum \citep{2017A&A...604A...6S} 
from earlier observations and found 
tentative detection of the north arc (Figure \ref{fig_colorimages}), confirming its reliability.
In conclusion, most, if not all, compact cores 
and extended structures in the continuum emission of the QUARKS survey can be reliably recovered by 
self-calibration in our data reduction.

\subsection{Spectral cube} 
The calibration tables   calculated from the self-calibration of continuum data
were applied to all the channels, including the emission line channels.
Subsequently,  polynomial fitting was conducted to the data of the line-free channels in the visibility domain, and
the continuum components were substracted using the {\it uvcontsub} task of CASA.
The visibility data of each SPW were then fed to CASA-{\it tclean} to generate cube of spectral lines. 
Thus, all the spectral cubes we generated are continuum-subtracted. 
The deconvolution algorithm of {\it tclean} was chosen as {\it multiscale}.
Since the dynamic ranges of the spectral images are overall lower than that of continuum images,
in practice, a uniform threshold of 15 mJy beam$^{-1}$ (corresponding to a brightness temperature, 
$T_{\rm b}$, threshold of 
3.8 K for a beam size of 0.3\arcsec) was chosen to optimize the performance of 
the cleaning of spectral cubes for most sources. As a result,
the rms of the spectral cube of the QUARKS survey 
is $\sim$3 mJy beam$^{-1}$ per channel, equivalent to  an rms of $T_{\rm b}$ of
0.8 K per channel.

\begin{figure*}[!thb]
\centering
\includegraphics[width=0.99\linewidth]{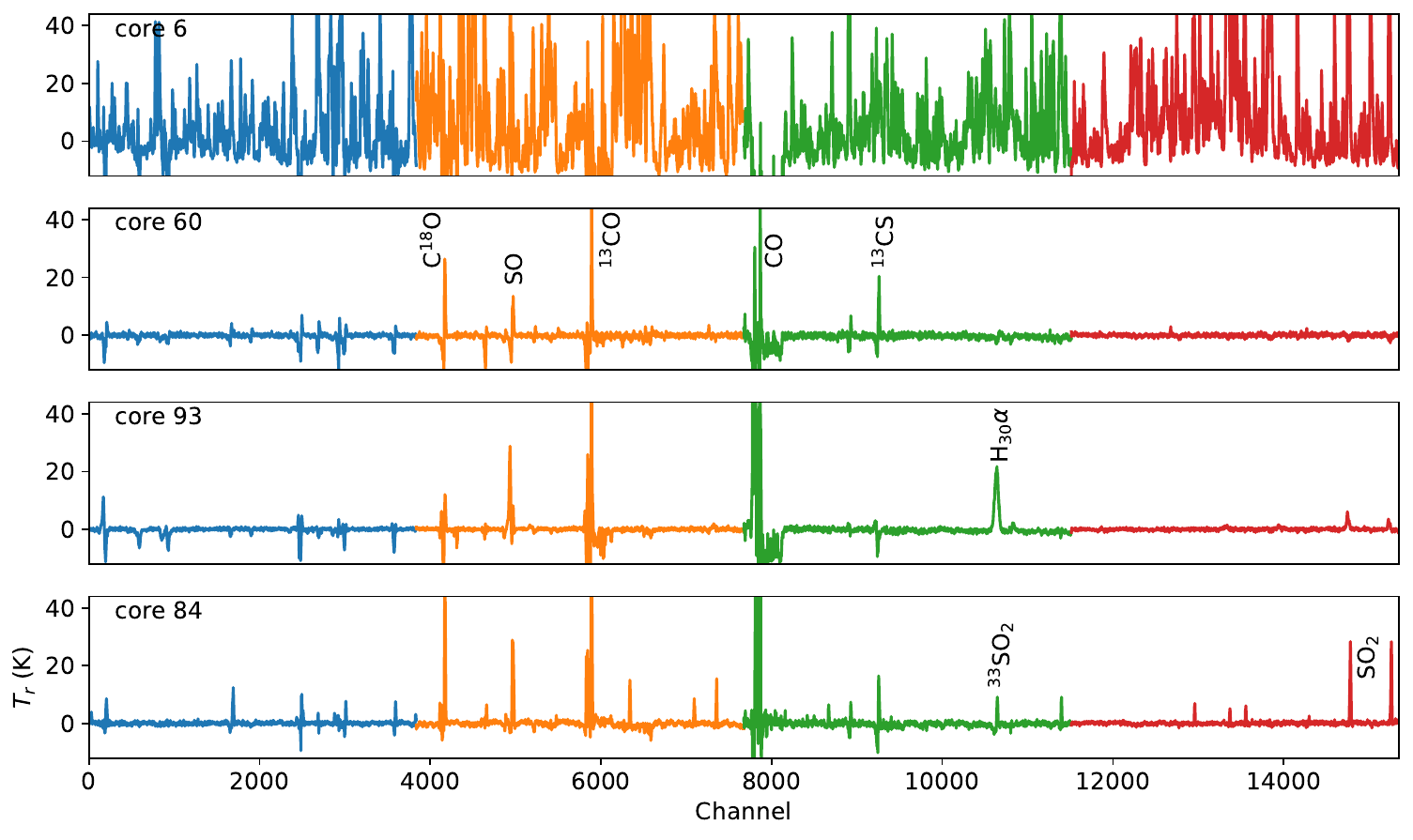}
\caption{Example spectra of cores of different types (Section \ref{sechotcores}). The blue, orange, green and red 
lines are the spectra of SPWs 1--4. See Figure \ref{fig_sgrb2cores} for the locations of these cores.
The spectra of hot core (core 6) is for comparison. 
The core 84 shows more than 20 lines of sulfur-bearing species. 
See Figures \ref{fig_core6spe}  for the zoom-in spectra of core 6.
\label{example_corespe} }
\end{figure*}

\section{A first look at Sgr B2(M)} \label{sec_sgrb2}
The giant molecular cloud Sgr B2 is 
the most massive \citep[$\sim 10^7$ $M_{\sun}$; e.g.,][]{1990ApJ...350..186G} region with ongoing high-mass star formation in the Galaxy.
It is an exceptional region in the Central Molecular Zone (CMZ), which appears
to be deficient in star formation \citep[e.g.,][]{1983A&A...117..343G,2017A&A...603A..89K}.
Its distance,  estimated from the newest version of the distance calculator of \citet{2016ApJ...823...77R}, is 8.3 kpc (Table \ref{tab_targets}), 
consistent with the values adopted by previous studies \citep[e.g.,][]{2018ApJ...853..171G,2019A&A...630A..73M}
based on the results of \citet{2014ApJ...783..130R}.
Sgr B2(M) is one of the two  most well-studied
`hot cores' (Sgr B2(M) and Sgr B2(N)) located in the central part of Sgr B2, which are actually high-mass protoclusters
containing hundreds of protostellar cores
\citep{2018ApJ...853..171G,2023A&A...676A.121M}.
Sgr B2 has also provided the  detection of most 
molecules discovered so far in the interstellar medium, and thus is considered  a very good target for 
studying astrochemistry \citep[e.g.,][]{2008A&A...482..179B,2022ApJS..259...30M}.
In this section, we explore the continuum (Section \ref{sec_srgb2cores}) 
and spectral (Section \ref{sechotcores}) data, and present some links between the data and 
science topics of the QUARKS survey through a brief case study of Sgr B2(M).  


\subsection{Mini-starburst in the protocluster Sgr B2(M) }\label{sec_srgb2cores}
First, we manually identify the cores and core-like substructures in the continuum map using the 
image viewer
CARTA\footnote{\url{https://cartavis.org/}}.
The criteria of a core are (1) it should have peak intensity larger than 5$\sigma$;
(2) it is  isolated or the major part of its emission region should 
not merge into its neighbouring cores.
In total, 97 cores were recognized.
We then fitted these cores using 2-D Gaussian functions,
using the optimization algorithm provided by the Python package, 
{\it lmfit}\footnote{\url{https://pypi.org/project/lmfit/}}. 
The catalog of the cores, including the 
position, FWHMs of the major ($L_{\rm maj}$) and minor ($L_{\rm min}$) axes, 
position angle (PA), peak intensity ($I_{\rm peak}$), and integrated flux
are listed in Table \ref{coretable}.
The 1.3 mm observation of \citet{2017A&A...604A...6S} revealed 27 cores in Sgr B2(M),
which are marked by the black crosses in panel (b) of Figure \ref{fig_colorimages}.
We crosschecked the two core catalogs, and the results are listed in the last column of Table \ref{coretable}.
All the cores of \citet{2017A&A...604A...6S}
can be identified in the QUARKS continuum image. 
The QUARKS survey has better angular resolution  and continuum sensitivity 
compared with that of \citet{2017A&A...604A...6S}, 
and thus our survey is capable of resolving many smaller-scale cores and identifying 
more weak entities.
For example, the cluster of protostars located in the southwest of the central part of Sgr B2(M),
which cannot be well resolved by \citet{2017A&A...604A...6S}, is revealed in our survey as
a collection of grape-like cores.
These cores may have evolved from a hub-filamentary structure,
but their intrinsic characteristics are to be further explored.

The fluxes of the cores identified in the QUARKS continuum image
range from 5 mJy to 1 Jy.
Assuming a uniform dust temperature of 100 K (the typical value of Sgr B2(M) by \citealt{2023A&A...676A.121M}) and a gas-to-dust mass ratio of 100,
this corresponds to a mass range of approximately 1 to 100 $M_{\sun}$.
For the brightest two cores (i.e., Cores 1 and 2 having $I_{\rm peak}\gtrsim 0.5$ Jy beam$^{-1}$), the optically-thin assumption could underestimate the core mass. For the remaining cores, their peak fluxes correspond to
peak brightness temperatures lower than 100 K, suggesting that their mass estimate could not be severely affected by that assumption.
The total mass of the cores in Sgr B2(M) is 2800 $M_{\sun}$,
yielding a mean mass of 30 $M_{\sun}$, much lower than the 
value of \citet{2017A&A...604A...6S} which is 150 $M_{\sun}$. 
This is expected since 
the QUARKS survey can probe smaller fragmented scales of Sgr B2(M) due to  higher sensitivity. 
The high-mass ($>$5 $M_{\sun}$) end of the core mass function (CMF)  show a power-law 
distribution ($d N/d \log(m)\propto m^{-\alpha}$) with a power index of $0.8\pm0.1$ (Figure \ref{fig_cmf}),
implying a top-heavy shape as compared to the canonical initial mass function \citep{2010ARA&A..48..339B,2019ApJ...886..102S,2022A&A...662A...8M}.  
The brightest cores may be even more massive 
if taking into account the effects of optical depths \citep{2017A&A...604A...6S},
which would result in a more top-heavy CMF. 
The kink in the high mass end result from the interference of dust shells of ionized regions
(e.g., Cores 92 and 93 in Figure \ref{fig_sgrb2cores}), which
are extended and thus have relative large fluxes. 
The emission of the shells at
1.3 mm is probably highly contaminated by free-free ionized gas
emission \citep{2017A&A...604A...6S}.
The CMF of Sgr B2(M) is similar to that of  Sgr B1-off, the mass range and power index of which are 
0.3--230 $M_{\sun}$ and $0.83\pm 0.21$, respectively \citep{2020ApJ...894L..14L}.
Note that Sgr B1-off is less evolved than Sgr B2(M), and a dust temperature of 20 K was adopted by 
\citet{2020ApJ...894L..14L}.
These results may imply that the evolutionary tracks of the
two regions may be regulated by some shared parameters of the environments,
although the varying dust temperatures provide challenges for precisely deriving the core masses.
On the other hand, the similarity of the CMFs  between Sgr B2(M) and other 
high-mass star forming regions \citep[with $\alpha<1$; e.g.,][]{2018ApJ...853..160C,2020ApJ...894L..14L,2023A&A...674A..76P} 
implies that the QUARKS is sensitive
enough to find out most of the high-mass cores in Sgr B2(M).
The core density of Sgr B2(M) is then derived to be $\sim 100$ pc$^{-2}$.
This value is much higher than the value (19 pc$^{-2}$) of
the  mini-starburst ridges of W43 \citep{2022A&A...664A..26P,2023A&A...674A..76P},
implying stronger mini-starburst activities in Sgr B2(M) compared with the W43
mini-starburst region, 
an analog of a forming super star cluster \citep{2010A&A...518L..90B}.

\subsection{COM-rich hot cores in Sgr B2(M)} \label{sechotcores}
Among the 97 cores, core 6 has the richest spectral line features (Figure \ref{fig_core6spe}).
The spectra of core 6 are shown in Figure \ref{fig_core6spe} as 
a gallery of the spectral lines that can be detected in the four SPWs of the QUARKS survey. 
Note that the spectra of core 6 do not contain all transitions that can be detected by the QUARKS survey.
Some important lines listed in Table 2 can not be seen in the spectra of core 6.
For example, no broad ($\Delta V>20$ km s$^{-1}$) line feature contributed by H$_{30}\alpha$ was detected.
A weak and narrow ($\Delta V\sim 6$ km s$^{-1}$) line feature can be seen at a frequency 
close to that of H$_{30}\alpha$, and
this line feature can be assigned to a transition of $^{33}$SO$_{2}$ through LTE fitting
as described in \citet{2022ApJS..263...13L}. It
implies that the core 6 is a `pure' hot core and does not harbor an UC H{\sc ii} region that is detectable by
the QUARKS survey based on the data of H30$\alpha$.
The emission of N$_2$D$^+$, a cold gas tracer, is also not detected in core 6, a COM-rich hot core.

We integrated the line groups of CH$_3$CN from 220.52 to 220.7 GHz 
and  NH$_2$CHO from 233.24 to 233.85 GHz (see Figure \ref{fig_core6spe}),
and the integrated intensity maps of the two line groups of Sgr B2(M)
are shown in Figure \ref{CH3CNfig}. 
CH$_3$CN and NH$_2$CHO are both COM species, and the two line groups trace COM-rich hot gas.
The emission of the two line groups is concentrated on 
a similar region, 
the central part of Sgr B2(M) with  a size of $\sim$8\arcsec~(0.3 pc).
Therefore, the central part of Sgr B2(M) can be treated as a single `big hot core'. The molecular gas inside and among individual dense cores within this `big hot core' is all  COM-rich hot gas.
The line group of NH$_2$CHO is more centrally peaked than that of CH$_3$CN,
implying different excitation states or chemical segregation within the `big hot core'.
The integrated intensity of the two line groups show another peak around core 41, 
which is more like a normal hot core heated by the
massive protostars within it.

More than half of the continuum cores are not located within the COM-rich emission regions (Figure \ref{CH3CNfig}). 
Those cores can be divided into three types according to their spectral features, including (1) line-poor cores
with no emission lines other than the  strong lines of common species such as 
CO and $^{13}$CS, (2) cores associated with H{\sc ii} regions showing strong emission of H30$\alpha$, and (3) sulfur-rich cores with very strong 
emission of sulfur-bearing species such as SO, SO$_2$, and their isotopologues.

We show the spectra of cores 60, 93, and 84 in Figure \ref{example_corespe} as examples of 
types 1, 2, and 3, respectively. Core 60 of this work is one of the 27 cores identified by \citet{2017A&A...604A...6S} in Sgr B2(M).
\citet{2017A&A...604A...6S} treated all those 27 cores as hot cores. However, 
some of these cores, such as their core A23 (core 60 in this work),
exhibit a deficiency in spectral lines. 
The dust temperatures of these 27 cores of \citet{2017A&A...604A...6S}  span a range of  
46--162 K,  as reported by \citet{2023A&A...676A.121M}, with core A23 having a dust temperature of 
72$\pm$38. 
Considering the diversity of dust temperatures, 
we classify cores with COM-rich line features
as hot cores, specifically those 
enclosed within the red contours in Figure \ref{CH3CNfig}. 
The sulfur-rich cores also tend to have high temperatures but are mainly located on the faint north arc
(Figure \ref{fig_sgrb2cores}), with a very different spatial distribution
compared with other gas components of Sgr B2(M) (Liu et al. in prep).

\section{Scientific objectives of the QUARKS survey} \label{sec_stopic}

Some scientific objectives (such as fragmentation, core mass function, astro-chemistry) of the QUARKS survey have been briefly mentioned above (Section \ref{sec_sgrb2}), through looking at an exemplar source, the Sgr B2(M). 
There are, however, more science topics that can be addressed by the QUARKS survey data but have not 
been fully discussed, as discussed below.

Hunting for massive starless cores is an important topic of this project. 
Whether or not massive starless cores exist serves as a key discriminator between star formation theories, 
but still in dispute  \citep[e.g.,][]{2003ApJ...585..850M,2004MNRAS.349..735B,2008ApJ...674..316H}.
However, hunting for high-mass starless cores has proven to be very challenging, and only a few promising
candidates have been reported \citep[e.g.,][]{2014ApJ...796L...2C,2014MNRAS.439.3275W,2017ApJ...834..193K}.
Most of the efforts in the search for massive starless cores have been focused on 
the early stages of massive clumps (e.g., infrared dark clouds).
However, observations indicated that gas accretion inside protoclusters 
(e.g., through hub-filamentary structures) would continue 
before being destroyed by stellar feedback 
\citep[e.g.,][]{2021ApJ...915L..10S,2022MNRAS.514.6038Z,2023arXiv230914684X}.
Several numerical simulations suggest that thermal feedback from OB protostars and the strong magnetic field
of proto-stellar clusters can play a crucial role in reducing the level of fragmentation 
and producing massive dense cores 
\citep[e.g.,][]{2007ApJ...656..959K,2009ApJ...704L.124O,2013ApJ...766...97M}.
Thus, it is possible that the newly assembled gas within evolving protoclusters  
could form massive starless cores.
During the pioneer exploration of the QUARKS data,
we have found a massive starless core candidate in  
I18507+0121, which shows no emission lines except for the line of N$_2$D$^+$ (Mai et al. in prep).  
The QUARKS survey can provide the 
high-mass star formation community additional samples of massive pre-stellar cores to explore.

How stellar feedback from outflows, H{\sc ii} regions and stellar winds from formed OB protostars influences the formation of new generation of stars and reshapes gas distribution in clouds is highly debated
\citep[e.g.,][]{2014prpl.conf..243K,2017MNRAS.466.3293P,2020PASP..132j4301S,2023MNRAS.520..322Z,2023A&A...674A..76P,2023MNRAS.521.5712H}.
The ATOMS survey studied the impact of UC H{\sc ii} regions (Zhang et al. in prep.) on their surrounding molecular gas,
but detailed investigation was limited by sensitivity, and resolution.
The high resolution of the QUARKS survey could help to study how the feedback of 
UC H{\sc ii} regions (traced by H$_{30}\alpha$)  influence
the surrounding star formation activities, e.g., the core masses under the impact of H{\sc ii} region on a scale of thousands au. The thermal feedback (e.g. shocks) from the ionized gas of the young H{\sc ii} regions could be reflected in the velocity and temperature distributions over the surrounding dense cores, which could be revealed by molecular transitions such as CH$_3$CN, CH$_3$OH and their isotopologues \citep{2018A&A...616A..66M} . The outflow tracers such as $^{12}$CO J=2-1, $^{13}$CO J=2-1 and SiO J=5-4 included in the QUARKS survey are very helpful for detecting molecular outflows and investigate the role of outflow feedback in maintain turbulence \citep[e.g.,][]{2020ApJ...903..119L,2021MNRAS.507.4316B} or heating gas \citep{2012ApJ...745L..30W} in gas clumps.

For protocluster within 3 kpc, 
a resolution of better than 1000 au would enable us to search for massive disks \citep{2016ApJ...823..125C} and large-scale ($>$10,000 au) gas streamers or spiral arms that may feed the disks \citep{2020NatAs...4.1158P,2023ApJ...953...82L} with various gas tracers (e.g., CH$_3$CN, SO).

Simulations and recent observational studies revealed time-dependent co-evolution between clump and core masses 
\citep[e.g.,][]{2004MNRAS.349..735B,2021MNRAS.508.2964A,2019ApJ...886..102S,2023arXiv230914684X}.
The QUARKS survey would help to reveal the co-evolution between protoclusters and protostars
in different environments.
Species such as N$_2$D$^+$, H$_2$CO, CH$_3$CN, and SO$_2$ are good tracers of 
gas temperature ranging from 10's
to hundreds Kelvin, enabling the QUARKS survey to investigate the evolution of 
cores  in different stages and other substructures found inside protoclusters. 
We expect that the QUARKS survey can help the study of the co-evolution of the multi-scale
structures of/within protoclusters, through case studies of typical sources, and, more
importantly, through unbiased statistical studies.  

\section{Link to other ALMA large survey programs}

High-resolution follow-up studies of the discoveries and investigations carried out in  the ATOMS survey \citep{2020MNRAS.496.2790L}
is an important goal of the QUARKS survey. 
How the protostars in protoclusters accumulate mass and 
in turn affect their natal clumps and regulate new star formation
is an important topic for both projects. The ATOMS survey has revealed ubiquitous filaments as well as filament-hub systems (HFSs) in protoclusters from dense core scales ($\sim$0.1 pc) to clump/cloud scales ($\sim$1-10 pc), and has found evidences for filamentary accretion at all scales \citep{2022MNRAS.514.6038Z}. 
The QUARKS survey can be used to check whether HFSs will further fragment down to a scale of 0.01 pc,
and can inspect how they feed individual protostars within protoclusters.
The QUARKS survey could also be used to establish the sample of hot cores.  
The ATOMS survey found  60 hot cores, 
contributing so far the largest uniform sample of hot cores observed with similar angular resolution and spectral 
coverage \citep{2022MNRAS.511.3463Q}. The QUARKS survey, with much higher resolution, will resolve these hot cores, and help to  statistically study the chemical diversities of COMs and the evolution of hot cores. 

Some large programs of ALMA,  e.g., the ALMA-IMF \citep{2022A&A...662A...8M} and ALMAGAL  (PI:
Sergio Molinari),
have science goals similar to or overlapping with that of the QUARKS survey.  
Compared with those programs, the  QUARKS survey has some unique advantages.
The ALMA-IMF observed 
15  extreme ($M_{\rm clump} > 2500$ M$_{\sun}$) and nearby (2--5.5 kpc) protoclusters in mosaic mode
at a  spatial resolution of $\sim$2000 au, aiming to
make breakthroughs in studying the origin of the initial mass function (IMF) of stars.  
The single pointing mode of the QUARKS survey
enables to generate quickly observe a relatively larger (139) and unbiased (Section \ref{sec_sample}) sample of protoculsters. 
Guided by the observations of the ATOMS survey in Band 3 \citep{2020MNRAS.496.2790L},
QUARKS focuses on the dense  kernels of the protoclusters that harbour majority of the massive protostars. 
In Sgr B2(M),  only 5 compact cores were detected by the ATOMS survey, and 
the number by the QUARKS survey  is more than 15 times that value.
In the weak-continuum source, IRAS 08448-4343, more than 20 dense cores have been revealed by the  QUARKS survey (Figure \ref{fig_sgrb2_compareselfcal}).
The ATOMS survey identified 453 compact cores, and it is reasonable to expect that
the QUARKS survey would detect at least 10 dense cores for each source on average and 1500 ones in total.
The total bandwidth (7.5 GHz)  of the QUARKS survey is twice the total bandwidth (3.75 GHz) of the ALMA-IMF.
The wide bandwidth makes the QUARKS survey ideal 
for studying the chemical properties of individual protostars
and of the protoculster as a whole,
e.g., investigating the chemical differentiation between O- and N-bearing COMs
\citep[e.g.,][]{1999ApJ...514L..43W,2022MNRAS.511.3463Q,2022MNRAS.512.4419P},
and exploring the formation mechanisms for individual complex organic molecules (e.g., CH$_3$COCH$_3$ and CH$_{3}$CHO, 
Shi et al. 2023, submitted).
The QUARKS survey can help to statistically study
the properties of protostars in an unbiased sample of protoclusters, and thus fully investigate the CMF at the high-mass end in widely different
Galactic environments and at different enolutionary stages,
bined by distances and protocluster properties.  

ALMAGAL observed more than 1000 dense clumps with 
$M>500$ $M_{\sun}$ and $d < 7.5$ kpc with a linear resolution of $\sim$1000 au
and a mass sensitivity of 0.3 $M_{\sun}$ at 1.3 mm, similar to the QUARKS survey. 
All the four SPWs of ALMAGAL are located within the frequency range of 216.9--221 GHz,
corresponding to the SPWs 1 and 2 of the QUARKS survey.
Emission lines of some important species, such as $^{12}$CO and N$_2$D$^+$, and strong recombination lines (the $\alpha$ lines of
hydrogen) were not covered by ALMAGAL.  
QUARKS covered the transtions of all the three isotologues of carbon monoxide (CO, $^{13}$CO and C$^{18}$O),
and in combination with the ATOMS survey, we can study the 
large-scale kinematics and structures up to 1 pc (30\arcsec~at a distance of 8 kpc). 
N$_2$D$^+$ is very helpful for searching massive starless cores \citep{2017ApJ...834..193K}.
Furthermore, the SPW4 of the QUARKS survey is free of strong lines, 
and it is very useful for searching and studying COMs in hot cores (Figure \ref{fig_core6spe}), 
while recombination lines could help identify H{\sc ii} regions.

\section{Summary}\label{secsummary}
We have presented an overview of the 
ALMA-QUARKS survey. 
The survey is observing 139 protoclusters, consisting of 156 pointings with the ALMA at Band 6. 
These protoclusters were selected as the dense kernels of massive clumps revealed by the 
ATOMS survey.
The primary goal of the QUARKS survey was to investigate the star formation process within protoclusters down to a scale of 1000 au 
in an unbiased and statistical manner. 
This paper has outlined the observations and data reduction of the QUARKS survey, and provides a first look at Sgr B2(M).
The QUARKS survey’s wide bandwidth (7.5 GHz) and high resolution ($\sim 0.3\arcsec$) allow us to resolve more compact cores than the ATOMS survey, 
and detect previously unrevealed fainter filamentary structures. 
The spectral windows cover transitions of species such as CO, SO, N$_2$D$^+$, SiO, H$_{30}\alpha$, 
and many complex organic molecules, tracing gas components with varying temperatures and spatial extents.
The unique observing setups of the QUARKS survey offer an opportunity to enhance our understanding of several scientific topics of star formation, 
including the mass transfer process inside protoclusters by hub-filamentary structures, the existence of massive starless cores, the physical and chemical 
properties of dense cores within protoclusters, and protocluster feedback. Preliminary analysis towards Sgr B2 (M2) indicates that our data reduction ensures
high-quality continuum and spectral line data for realizing these scientific objectives. Specifically, the self-calibration approach can further improve data quality 
particularly for targeted sources with strong emission.

\normalem
\begin{acknowledgements}
{\small
This work has been supported by the National Key R\&D Program
of China (No. 2022YFA1603100), National Natural Science Foundation of China (NSFC) through grants No. 12203086, 12033005, 12073061, 12122307, and 12103045. Moreover, X.L. has also been supported by CPSF No. 2022M723278; T.L by the international partnership program of Chinese Academy of Sciences through grant No.114231KYSB20200009, Shanghai Pujiang Program 20PJ1415500,  the science research grants from the China Manned Space Project with no. CMS-CSST-2021-B06; H.L. by Yunnan Fundamental Research Project (grant No.\,202301AT070118).
This research was carried out in part at the Jet Propulsion Laboratory, which is operated by the California Institute of Technology under a contract with the National Aeronautics
and Space Administration (80NM0018D0004). 
AS, GG, LB, and DM acknowledge support from the ANID BASAL project FB210003. 
AS also acknowledges support from the Fondecyt Regular (project code 1220610).
PS was partially supported by a Grant-in-Aid for Scientific Research (KAKENHI Number JP22H01271 and JP23H01221) of JSPS. 
K.T. was supported by JSPS KAKENHI (Grant Number JP20H05645). 
This paper makes use of the following ALMA data: ADS/JAO.ALMA\#2019.1.00685.S, 2021.1.00095.S and 2023.1.00425. ALMA is a partnership of ESO (representing its member states), NSF (USA) and NINS (Japan), together with NRC (Canada), MOST and ASIAA (Taiwan), and KASI (Republic of Korea), in cooperation with the Republic of Chile. The Joint ALMA Observatory is operated by ESO, AUI/NRAO and NAOJ.
We show warm thanks to the anonymous referee for providing helpful comments for improving the paper.
}
\end{acknowledgements}
  
\bibliographystyle{raa}
\bibliography{QUARKS_paperI}

\clearpage
\appendix
\section{Source catalogue}
Table \ref{tab_targets} lists all the observed targets of the QUARKS survey,
including the target names, coordinates, velocity ($V_{\rm LSR}$), and distance ($D$).
For velocity and distance, both of the values adopted by the ATOMS survey 
(marked by a superscript of `old') and the QUARKS survey (marked by a superscript of `new') are listed.
Table \ref{coretable} lists the continuum cores of Sgr B2(M) revealed by the QUARKS survey.

\begin{table*}[!thb]
\centering
\caption{The targets of the QUARKS survey$^{(1)}$. \label{tab_targets}}
\begin{tabular}{llcccccccc}
\hline\hline
ID & Target$^{\ (2)}$ & RA (J2000) & DEC (J2000) &Glon&Glat& $V_{\rm LSR}^{\rm old (3)}$ &$V_{\rm LSR}^{\rm new}$ & $D^{\rm old}$ & $D^{\rm new}$\\
   &        &    &     & ($\degr$)    & ($\degr$)   & (km s$^{-1}$) & (km s$^{-1}$)& (kpc)& (kpc)\\
\hline
1 & I08303-4303 & 08:32:09.190 & -43:13:44.30 & 261.64503 & -2.08742 & 14.3 & 14.5 & 2.3 & 2.35 \\
2 & I08448-4343\_1 & 08:46:32.270 & -43:54:35.70 & 263.77191 & -0.43510 & 3.7 & 2.8 & 0.7 & 1.22 \\
3 & I08448-4343\_2 & 08:46:34.340 & -43:54:30.50 & 263.77467 & -0.42935 & 3.7 & 2.8 & 0.7 & 1.22 \\
4 & I08470-4243 & 08:48:47.720 & -42:54:22.00 & 263.24857 & +0.51455 & 12.0 & 12.1 & 2.1 & 2.27 \\
5 & I09002-4732 & 09:01:54.140 & -47:44:09.30 & 268.42188 & -0.84936 & 3.1 & 3.2 & 1.2 & 1.62 \\
6 & I09018-4816 & 09:03:32.820 & -48:28:06.20 & 269.15212 & -1.13009 & 10.3 & 9.9 & 2.6 & 2.57 \\
7 & I09094-4803 & 09:11:08.260 & -48:15:54.60 & 269.85280 & -0.06249 & 74.6 & 74.4 & 9.6 & 8.75 \\
8 & I10365-5803\_1 & 10:38:32.070 & -58:19:01.10 & 286.20619 & +0.17153 & -19.0 & -20.1 & 2.4 & 2.49 \\
9 & I10365-5803\_2 & 10:38:32.990 & -58:19:16.70 & 286.21007 & +0.16873 & -19.0 & -20.1 & 2.4 & 2.5 \\
10 & I11298-6155 & 11:32:06.030 & -62:12:20.50 & 293.82778 & -0.74454 & 32.9 & 33.4 & 10.0 & 9.6 \\
11 & I11332-6258 & 11:35:32.230 & -63:14:46.80 & 294.51178 & -1.62216 & -15.4 & -15.4 & 1.9 & 1.4 \\
12 & I12320-6122 & 12:34:53.380 & -61:39:46.90 & 300.96916 & +1.14564 & -42.5 & -43.2 & 3.43 & 4.17 \\
13 & I12326-6245 & 12:35:34.810 & -63:02:32.10 & 301.13583 & -0.22582 & -39.6 & -39.5 & 4.61 & 4.21 \\
14 & I12383-6128 & 12:41:17.320 & -61:44:38.60 & 301.73081 & +1.10402 & -39.1 & -39.4 & 3.27 & 4.12 \\
15 & I12572-6316\_1 & 13:00:24.030 & -63:32:31.90 & 303.93008 & -0.68782 & 30.9 & 30.4 & 11.57 & 11.63 \\
16 & I12572-6316\_2 & 13:00:28.730 & -63:32:37.30 & 303.93875 & -0.68962 & 30.9 & 30.4 & 11.57 & 11.63 \\
17 & I13079-6218\_1 & 13:11:13.730 & -62:34:40.20 & 305.20834 & +0.20641 & -42.6 & -41.3 & 3.8 & 3.11 \\
18 & I13079-6218\_2 & 13:11:09.500 & -62:34:39.70 & 305.20026 & +0.20717 & -42.6 & -41.3 & 3.8 & 3.11 \\
19 & I13080-6229 & 13:11:14.280 & -62:44:58.30 & 305.19621 & +0.03514 & -35.6 & -35.7 & 3.8 & 2.68 \\
20 & I13111-6228 & 13:14:26.490 & -62:44:28.30 & 305.56239 & +0.01301 & -38.8 & -39.5 & 3.8 & 2.97 \\
21 & I13134-6242 & 13:16:42.990 & -62:58:29.30 & 305.79886 & -0.24386 & -31.5 & -32.0 & 3.8 & 4.93 \\
22 & I13140-6226 & 13:17:15.900 & -62:42:27.00 & 305.88761 & +0.01594 & -33.9 & -34.4 & 3.8 & 4.88 \\
23 & I13291-6229\_1 & 13:32:31.770 & -62:45:11.80 & 307.61519 & -0.25577 & -36.5 & -37.6 & 2.9 & 2.66 \\
24 & I13291-6229\_2 & 13:32:34.580 & -62:45:27.00 & 307.61981 & -0.26079 & -36.5 & -37.6 & 2.9 & 2.66 \\
25 & I13291-6249 & 13:32:31.230 & -63:05:21.80 & 307.56083 & -0.58746 & -34.7 & -33.4 & 7.61 & 7.72 \\
26 & I13295-6152 & 13:32:53.490 & -62:07:49.30 & 307.75582 & +0.35251 & -44.4 & -44.7 & 3.89 & 3.27 \\
27 & I13471-6120 & 13:50:42.100 & -61:35:14.90 & 309.92084 & +0.47748 & -56.7 & -57.8 & 5.46 & 5.17 \\
28 & I13484-6100 & 13:51:58.640 & -61:15:43.30 & 310.14416 & +0.75912 & -55.0 & -55.3 & 5.4 & 6.64 \\
29 & I14013-6105 & 14:04:54.560 & -61:20:10.70 & 311.62568 & +0.28912 & -48.1 & -55.4 & 4.12 & 3.9 \\
30 & I14050-6056 & 14:08:42.150 & -61:10:43.00 & 312.10777 & +0.30909 & -47.1 & -47.6 & 3.42 & 3.19 \\
31 & I14164-6028 & 14:20:08.230 & -60:42:05.00 & 313.57582 & +0.32418 & -46.5 & -46.7 & 3.19 & 2.95 \\
32 & I14206-6151 & 14:24:22.810 & -62:05:22.70 & 313.57583 & -1.15400 & -50.0 & -49.7 & 3.29 & 3.19 \\
33 & I14212-6131\_1 & 14:25:00.980 & -61:44:57.50 & 313.76582 & -0.86186 & -50.5 & -50.8 & 3.44 & 3.29 \\
34 & I14212-6131\_2 & 14:25:04.240 & -61:44:46.40 & 313.77292 & -0.86125 & -50.5 & -50.8 & 3.44 & 3.29 \\
35 & I14382-6017 & 14:42:01.900 & -60:30:22.50 & 316.13901 & -0.50189 & -60.7 & -60.7 & 7.69 & 7.75 \\
36 & I14498-5856 & 14:53:42.530 & -59:08:53.20 & 318.04937 & +0.08691 & -49.3 & -50.2 & 3.16 & 2.99 \\
37 & I15122-5801 & 15:16:05.650 & -58:11:39.60 & 321.05251 & -0.50580 & -60.9 & -60.9 & 9.26 & 8.73 \\
38 & I15254-5621 & 15:29:19.480 & -56:31:23.20 & 323.45906 & -0.07914 & -67.3 & -68.6 & 4.0 & 4.42 \\
39 & I15290-5546 & 15:32:53.230 & -55:56:09.60 & 324.20074 & +0.12008 & -87.5 & -88.3 & 6.76 & 7.94 \\
40 & I15384-5348 & 15:42:17.510 & -53:58:28.20 & 326.44810 & +0.90660 & -41.0 & -41.1 & 1.82 & 2.41 \\
41 & I15394-5358 & 15:43:16.590 & -54:07:14.70 & 326.47465 & +0.70271 & -41.6 & -40.5 & 1.82 & 2.38 \\
42 & I15411-5352 & 15:44:59.460 & -54:02:14.30 & 326.72494 & +0.61576 & -41.5 & -41.8 & 1.82 & 2.41 \\
43 & I15437-5343 & 15:47:33.110 & -53:52:43.90 & 327.11942 & +0.50904 & -83.0 & -83.4 & 4.98 & 5.16 \\
44 & I15439-5449 & 15:47:49.820 & -54:58:28.10 & 326.47310 & -0.37656 & -54.6 & -54.8 & 3.29 & 3.4 \\
45 & I15502-5302 & 15:54:06.430 & -53:11:38.40 & 328.30750 & +0.43085 & -91.4 & -92.4 & 5.8 & 5.49 \\
46 & I15520-5234 & 15:55:48.390 & -52:43:09.80 & 328.80764 & +0.63243 & -41.3 & -41.8 & 2.65 & 2.56 \\
47 & I15522-5411\_1 & 15:56:06.950 & -54:19:58.90 & 327.80753 & -0.63475 & -46.7 & -47.5 & 2.73 & 2.89 \\
48 & I15522-5411\_2 & 15:56:08.220 & -54:19:35.70 & 327.81403 & -0.63179 & -46.7 & -47.5 & 2.73 & 2.89 \\
49 & I15557-5215 & 15:59:40.760 & -52:23:27.70 & 329.46907 & +0.50244 & -67.6 & -68.5 & 4.03 & 6.91 \\
50 & I15567-5236 & 16:00:32.860 & -52:44:45.10 & 329.33743 & +0.14749 & -107.1 & -107.5 & 5.99 & 5.22 \\
51 & I15570-5227 & 16:00:55.560 & -52:36:25.20 & 329.47166 & +0.21497 & -101.5 & -100.7 & 5.99 & 5.64 \\
52 & I15584-5247 & 16:02:19.630 & -52:55:28.40 & 329.42299 & -0.16370 & -76.8 & -76.4 & 4.41 & 4.87 \\
53 & I15596-5301 & 16:03:32.290 & -53:09:28.10 & 329.40576 & -0.45907 & -72.1 & -73.7 & 10.11 & 9.2 \\
54 & I16026-5035\_1 & 16:06:25.630 & -50:43:18.10 & 331.35920 & +1.06314 & -78.3 & -79.0 & 4.53 & 5.05 \\
55 & I16026-5035\_2 & 16:06:23.110 & -50:43:26.10 & 331.35277 & +1.06593 & -78.3 & -79.0 & 4.53 & 5.05 \\
\hline
\end{tabular}{\flushleft
$^{(1)}$ This table contains 156 rows.\\
$^{(2)}$ The two pointing targets associated with a same source were denoted by suffixes  ``\_1'' and ``\_2''.\\
$^{(3)}$ The $V_{\rm LSR}$ is the velocity revealed by surveys before the ATOMS survey. The $V_{\rm LSR}^{\rm new}$ 
is the velocity revealed by the H$^{13}$CO$^+$ data of the ATOMS survey. 
The $D^{\rm old}$ is the distance adopted by the ATOMS survey,
originally derived from the Galactic rotation curve by \citet{1996A&AS..115...81B}.
The $D^{\rm new}$ is derived from the coordinate and $V_{\rm LSR}^{\rm new}$, using the distance calculator of \citet{2016ApJ...823...77R}. See Section \ref{sec_sample} for details.\\
}
\end{table*}

\addtocounter{table}{-1}

\begin{table*}[!thb]
\centering
\caption{continued.}
\begin{tabular}{llcccccccc}
\hline\hline
ID & Target  & RA (J2000) & DEC (J2000) &Glon&Glat& $V_{\rm LSR}^{\rm old}$ &$V_{\rm LSR}^{\rm new}$ & $D^{\rm old}$ & $D^{\rm new}$\\
   &        &    &     & ($\degr$)    & ($\degr$)   & (km s$^{-1}$) & (km s$^{-1}$)& (kpc)& (kpc)\\
\hline
56 & I16037-5223 & 16:07:38.100 & -52:31:00.20 & 330.29433 & -0.39406 & -80.0 & -81.3 & 9.84 & 9.04 \\
57 & I16060-5146 & 16:09:52.850 & -51:54:54.70 & 330.95419 & -0.18248 & -91.6 & -92.1 & 5.3 & 5.39 \\
58 & I16065-5158 & 16:10:19.600 & -52:06:07.10 & 330.87788 & -0.36624 & -63.3 & -62.5 & 3.98 & 4.33 \\
59 & I16071-5142 & 16:10:59.010 & -51:50:21.60 & 331.13062 & -0.24240 & -87.0 & -86.5 & 5.3 & 5.29 \\
60 & I16076-5134 & 16:11:27.120 & -51:41:56.90 & 331.27918 & -0.18915 & -87.7 & -87.8 & 5.3 & 5.31 \\
61 & I16119-5048 & 16:15:45.650 & -50:55:58.50 & 332.29482 & -0.09513 & -48.2 & -48.6 & 3.1 & 3.42 \\
62 & I16132-5039\_1 & 16:17:01.920 & -50:46:51.00 & 332.54480 & -0.12485 & -47.5 & -47.6 & 3.1 & 3.36 \\
63 & I16132-5039\_2 & 16:17:02.690 & -50:47:07.10 & 332.54315 & -0.12947 & -47.5 & -47.6 & 3.1 & 3.36 \\
64 & I16164-5046 & 16:20:10.910 & -50:53:15.50 & 332.82560 & -0.54900 & -57.3 & -56.7 & 3.57 & 4.02 \\
65 & I16172-5028\_1 & 16:21:02.930 & -50:35:11.60 & 333.13502 & -0.43180 & -51.9 & -53.3 & 3.57 & 3.78 \\
66 & I16172-5028\_2 & 16:21:00.160 & -50:35:09.10 & 333.13032 & -0.42614 & -51.9 & -53.3 & 3.57 & 3.79 \\
67 & I16177-5018 & 16:21:31.490 & -50:25:04.50 & 333.30767 & -0.36574 & -50.2 & -50.0 & 3.57 & 3.53 \\
68 & I16272-4837 & 16:30:58.020 & -48:43:46.60 & 335.58505 & -0.28737 & -46.6 & -46.8 & 2.92 & 3.54 \\
69 & I16297-4757 & 16:33:29.290 & -48:03:31.50 & 336.36099 & -0.13529 & -79.6 & -79.3 & 5.03 & 5.1 \\
70 & I16304-4710 & 16:34:04.800 & -47:16:32.00 & 337.00368 & +0.32327 & -62.8 & -62.3 & 11.32 & 10.5 \\
71 & I16313-4729 & 16:34:55.440 & -47:35:40.90 & 336.86551 & +0.00213 & -73.7 & -74.1 & 4.71 & 4.99 \\
72 & I16318-4724 & 16:35:33.200 & -47:31:11.30 & 336.99245 & -0.02558 & -119.8 & -120.8 & 7.68 & 7.95 \\
73 & I16330-4725 & 16:36:42.600 & -47:31:30.80 & 337.11993 & -0.17360 & -75.1 & -75.4 & 10.99 & 9.96 \\
74 & I16344-4658 & 16:38:10.380 & -47:04:56.70 & 337.61479 & -0.06089 & -49.5 & -49.7 & 12.09 & 11.23 \\
75 & I16348-4654 & 16:38:29.420 & -47:00:39.70 & 337.70403 & -0.05342 & -46.5 & -47.8 & 12.09 & 11.33 \\
76 & I16351-4722 & 16:38:50.610 & -47:27:59.70 & 337.40474 & -0.40206 & -41.4 & -40.8 & 3.02 & 2.33 \\
77 & I16362-4639 & 16:39:57.700 & -46:45:04.30 & 338.06530 & -0.06792 & -38.8 & -38.6 & 3.01 & 2.26 \\
78 & I16372-4545 & 16:40:54.770 & -45:50:53.60 & 338.85013 & +0.40789 & -57.3 & -57.5 & 4.16 & 4.42 \\
79 & I16385-4619 & 16:42:13.980 & -46:25:27.20 & 338.56865 & -0.14425 & -117.0 & -117.0 & 7.11 & 5.76 \\
80 & I16424-4531 & 16:46:06.610 & -45:36:46.60 & 339.62257 & -0.12249 & -34.2 & -34.5 & 2.63 & 2.19 \\
81 & I16445-4459 & 16:48:05.180 & -45:05:08.60 & 340.24917 & -0.04582 & -121.3 & -122.2 & 7.95 & 7.72 \\
82 & I16458-4512 & 16:49:30.410 & -45:17:53.60 & 340.24749 & -0.37413 & -50.4 & -50.9 & 3.56 & 4.19 \\
83 & I16484-4603 & 16:52:03.990 & -46:08:24.60 & 339.88482 & -1.25558 & -32.0 & -32.4 & 2.1 & 2.17 \\
84 & I16487-4423\_1 & 16:52:23.670 & -44:27:52.30 & 341.21578 & -0.23579 & -43.4 & -44.0 & 3.26 & 2.65 \\
85 & I16487-4423\_2 & 16:52:21.590 & -44:28:00.60 & 341.21007 & -0.23248 & -43.4 & -44.0 & 3.26 & 2.65 \\
86 & I16489-4431 & 16:52:34.080 & -44:36:16.70 & 341.12713 & -0.34861 & -41.3 & -41.1 & 3.26 & 2.48 \\
87 & I16524-4300 & 16:56:03.060 & -43:04:47.30 & 342.70669 & +0.12514 & -40.8 & -40.6 & 3.43 & 2.52 \\
88 & I16547-4247 & 16:58:17.260 & -42:52:04.50 & 343.12750 & -0.06248 & -30.4 & -30.6 & 2.74 & 2.19 \\
89 & I16562-3959 & 16:59:41.450 & -40:03:34.60 & 345.49528 & +1.47062 & -12.6 & -11.3 & 2.38 & 1.37 \\
90 & I16571-4029 & 17:00:32.380 & -40:34:13.80 & 345.19280 & +1.02820 & -15.0 & -15.4 & 2.38 & 2.13 \\
91 & I17006-4215 & 17:04:13.200 & -42:19:54.20 & 344.22119 & -0.59458 & -23.2 & -24.3 & 2.21 & 2.85 \\
92 & I17008-4040 & 17:04:23.200 & -40:44:24.90 & 345.50460 & +0.34705 & -17.0 & -17.3 & 2.38 & 1.33 \\
93 & I17016-4124 & 17:05:11.020 & -41:29:07.80 & 345.00285 & -0.22408 & -27.1 & -26.8 & 1.37 & 3.16 \\
94 & I17136-3617 & 17:17:02.290 & -36:21:02.10 & 350.50570 & +0.95672 & -10.6 & -11.6 & 1.34 & 1.33 \\
95 & I17143-3700 & 17:17:45.650 & -37:03:11.80 & 350.01583 & +0.43248 & -31.1 & -31.7 & 12.67 & 11.71 \\
96 & I17158-3901 & 17:19:20.340 & -39:03:53.30 & 348.54925 & -0.97890 & -15.2 & -16.6 & 3.38 & 3.33 \\
97 & I17160-3707 & 17:19:27.310 & -37:11:00.40 & 350.10365 & +0.08132 & -69.5 & -69.4 & 10.53 & 10.28 \\
98 & I17175-3544 & 17:20:53.570 & -35:46:59.80 & 351.41758 & +0.64492 & -5.7 & -8.7 & 1.34 & 1.32 \\
99 & I17204-3636 & 17:23:50.320 & -36:38:58.10 & 351.04114 & -0.33570 & -18.2 & -18.1 & 3.32 & 3.27 \\
100 & I17220-3609 & 17:25:24.990 & -36:12:41.10 & 351.58185 & -0.35181 & -93.7 & -97.2 & 8.01 & 8.05 \\
101 & I17233-3606 & 17:26:42.800 & -36:09:16.80 & 351.77525 & -0.53699 & -2.7 & -3.1 & 1.34 & 1.32 \\
102 & I17244-3536 & 17:27:48.710 & -35:39:10.80 & 352.31584 & -0.44245 & -10.2 & -10.8 & 1.36 & 2.63 \\
103 & I17258-3637 & 17:29:17.130 & -36:40:12.60 & 351.63374 & -1.25360 & -11.9 & -10.5 & 2.59 & 1.33 \\
104 & I17269-3312\_1 & 17:30:15.200 & -33:14:57.50 & 354.59672 & +0.46787 & -21.0 & -21.4 & 4.38 & 4.53 \\
105 & I17269-3312\_2 & 17:30:14.120 & -33:14:22.60 & 354.60274 & +0.47635 & -21.0 & -21.4 & 4.38 & 4.53 \\
106 & I17271-3439\_1 & 17:30:26.210 & -34:41:45.40 & 353.40990 & -0.36021 & -18.2 & -17.2 & 3.1 & 3.66 \\
107 & I17271-3439\_2 & 17:30:27.710 & -34:41:45.90 & 353.41261 & -0.36457 & -18.2 & -17.2 & 3.1 & 3.67 \\
108 & I17278-3541 & 17:31:14.070 & -35:44:04.30 & 352.63152 & -1.06690 & -0.4 & -0.5 & 1.33 & 1.32 \\
109 & I17439-2845\_1 & 17:47:09.060 & -28:46:17.80 & 0.31480 & -0.20090 & 18.7 & 18.4 & 8.0 & 6.93 \\
110 & I17439-2845\_2 & 17:47:07.190 & -28:46:01.30 & 0.31518 & -0.19269 & 18.7 & 18.4 & 8.0 & 6.94 \\
111 & I17441-2822 & 17:47:19.790 & -28:23:05.70 & 0.66589 & -0.03412 & 50.8 & 50.8 & 8.1 & 8.3 \\
112 & I17455-2800 & 17:48:41.630 & -28:01:44.60 & 1.12584 & -0.10749 & -15.6 & -14.8 & 10.0 & 11.14 \\
113 & I17545-2357 & 17:57:34.490 & -23:58:04.30 & 5.63735 & +0.23748 & 7.9 & 8.8 & 2.93 & 3.0 \\
114 & I17589-2312 & 18:01:58.000 & -23:12:32.00 & 6.79633 & -0.25777 & 21.3 & 20.5 & 2.97 & 3.44 \\
115 & I17599-2148 & 18:03:01.470 & -21:48:06.30 & 8.14078 & +0.22409 & 18.6 & 18.6 & 2.99 & 3.44 \\
116 & I18032-2032 & 18:06:14.300 & -20:31:35.00 & 9.61969 & +0.19657 & 4.3 & 4.4 & 5.15 & 4.8 \\
117 & I18056-1952 & 18:08:38.180 & -19:51:49.00 & 10.47249 & +0.02752 & 66.7 & 66.4 & 8.55 & 8.44 \\
118 & I18075-2040 & 18:10:34.500 & -20:39:16.10 & 10.00023 & -0.75297 & 31.5 & 31.2 & 3.08 & 2.8 \\
119 & I18079-1756 & 18:10:50.770 & -17:55:45.20 & 12.41844 & +0.50544 & 18.0 & 17.8 & 1.83 & 2.32 \\
120 & I18089-1732 & 18:11:51.060 & -17:31:27.20 & 12.88863 & +0.49074 & 33.5 & 32.9 & 2.5 & 3.64 \\
\hline
\end{tabular}
\end{table*}

\addtocounter{table}{-1}

\begin{table*}[!thb]
\centering
\caption{continued.}
\begin{tabular}{llcccccccc}
\hline\hline
ID & Target  & RA (J2000) & DEC (J2000) &Glon&Glat& $V_{\rm LSR}^{\rm old}$ &$V_{\rm LSR}^{\rm new}$ & $D^{\rm old}$ & $D^{\rm new}$\\
   &        &    &     & ($\degr$)    & ($\degr$)   & (km s$^{-1}$) & (km s$^{-1}$)& (kpc)& (kpc)\\
\hline
121 & I18110-1854 & 18:14:01.050 & -18:53:23.30 & 11.93713 & -0.61570 & 37.0 & 38.5 & 3.37 & 3.17 \\
122 & I18116-1646 & 18:14:35.960 & -16:45:36.80 & 13.87386 & +0.28067 & 48.5 & 48.8 & 3.94 & 3.86 \\
123 & I18117-1753 & 18:14:39.250 & -17:51:59.80 & 12.90811 & -0.25932 & 36.7 & 37.1 & 2.57 & 3.63 \\
124 & I18134-1942 & 18:16:22.120 & -19:41:27.00 & 11.49742 & -1.48527 & 10.6 & 10.5 & 1.25 & 1.25 \\
125 & I18139-1842 & 18:16:51.590 & -18:41:34.70 & 12.43083 & -1.11328 & 39.8 & 39.8 & 3.02 & 3.0 \\
126 & I18159-1648 & 18:18:54.340 & -16:47:45.90 & 14.33225 & -0.64324 & 22.0 & 22.4 & 1.48 & 2.83 \\
127 & I18182-1433 & 18:21:09.220 & -14:31:46.80 & 16.58584 & -0.05082 & 59.1 & 59.3 & 4.71 & 3.66 \\
128 & I18223-1243 & 18:25:10.580 & -12:42:22.20 & 18.65473 & -0.05887 & 44.8 & 45.0 & 3.37 & 3.46 \\
129 & I18228-1312 & 18:25:42.270 & -13:10:17.00 & 18.30343 & -0.38990 & 32.3 & 32.8 & 3.21 & 3.2 \\
130 & I18236-1205 & 18:26:25.650 & -12:03:57.60 & 19.36348 & -0.03031 & 25.9 & 26.1 & 2.17 & 2.88 \\
131 & I18264-1152 & 18:29:14.500 & -11:50:19.50 & 19.88476 & -0.53396 & 43.2 & 43.6 & 3.33 & 3.35 \\
132 & I18290-0924\_1 & 18:31:42.980 & -09:22:26.00 & 22.35097 & +0.06876 & 84.2 & 84.0 & 5.34 & 4.69 \\
133 & I18290-0924\_2 & 18:31:43.740 & -09:22:19.00 & 22.35413 & +0.06689 & 84.2 & 84.0 & 5.34 & 4.69 \\
134 & I18308-0503\_1 & 18:33:29.780 & -05:01:04.30 & 26.41961 & +1.68770 & 42.9 & 42.6 & 3.1 & 3.75 \\
135 & I18308-0503\_2 & 18:33:30.850 & -05:00:58.70 & 26.42303 & +1.68447 & 42.9 & 42.6 & 3.1 & 3.75 \\
136 & I18311-0809 & 18:33:53.430 & -08:07:14.40 & 23.71065 & +0.17082 & 113.0 & 112.9 & 6.06 & 6.27 \\
137 & I18314-0720\_1 & 18:34:10.010 & -07:18:00.30 & 24.47044 & +0.48830 & 101.5 & 101.3 & 5.82 & 6.15 \\
138 & I18314-0720\_2 & 18:34:09.050 & -07:17:50.80 & 24.47095 & +0.49303 & 101.5 & 101.3 & 5.82 & 6.15 \\
139 & I18316-0602 & 18:34:20.580 & -05:59:41.60 & 25.64915 & +1.05073 & 42.8 & 42.7 & 2.09 & 3.76 \\
140 & I18317-0513 & 18:34:25.800 & -05:10:53.50 & 26.38134 & +1.40596 & 42.1 & 41.8 & 2.18 & 3.73 \\
141 & I18317-0757 & 18:34:24.900 & -07:54:47.60 & 23.95460 & +0.15119 & 80.7 & 81.2 & 4.79 & 4.7 \\
142 & I18341-0727 & 18:36:50.400 & -07:24:48.00 & 24.67487 & -0.15227 & 112.7 & 113.1 & 6.04 & 6.39 \\
143 & I18411-0338 & 18:43:46.260 & -03:35:23.90 & 28.86343 & +0.06630 & 102.8 & 103.5 & 7.41 & 5.46 \\
144 & I18434-0242 & 18:46:04.200 & -02:39:18.30 & 29.95723 & -0.01729 & 97.2 & 97.5 & 5.16 & 4.76 \\
145 & I18440-0148 & 18:46:36.390 & -01:45:22.00 & 30.81828 & +0.27376 & 97.5 & 97.6 & 5.16 & 6.04 \\
146 & I18445-0222 & 18:47:09.760 & -02:18:47.60 & 30.38596 & -0.10410 & 86.9 & 86.6 & 5.16 & 6.24 \\
147 & I18461-0113 & 18:48:42.230 & -01:10:05.40 & 31.58061 & +0.07556 & 96.1 & 96.0 & 5.16 & 5.67 \\
148 & I18469-0132 & 18:49:33.150 & -01:29:06.20 & 31.39533 & -0.25773 & 87.0 & 86.6 & 5.16 & 4.91 \\
149 & I18479-0005 & 18:50:31.200 & -00:01:56.00 & 32.79860 & +0.18954 & 14.6 & 14.6 & 12.96 & 12.87 \\
150 & I18507+0110 & 18:53:18.120 & +01:15:00.10 & 34.25691 & +0.15530 & 57.2 & 58.2 & 1.56 & 3.23 \\
151 & I18507+0121 & 18:53:18.150 & +01:25:22.40 & 34.41081 & +0.23402 & 57.9 & 57.8 & 1.56 & 3.09 \\
152 & I18517+0437 & 18:54:14.110 & +04:41:43.10 & 37.43000 & +1.51849 & 43.9 & 43.9 & 2.36 & 2.36 \\
153 & I18530+0215 & 18:55:33.610 & +02:19:09.00 & 35.46573 & +0.14083 & 74.1 & 76.9 & 4.67 & 5.08 \\
154 & I19078+0901 & 19:10:13.410 & +09:06:10.40 & 43.16581 & +0.01086 & 2.9 & 6.2 & 11.11 & 11.49 \\
155 & I19095+0930 & 19:11:53.900 & +09:35:45.90 & 43.79414 & -0.12749 & 43.7 & 43.8 & 6.02 & 9.07 \\
156 & I19097+0847 & 19:12:08.900 & +08:52:07.80 & 43.17804 & -0.51881 & 58.0 & 58.2 & 8.47 & 7.83 \\
\hline
\end{tabular}
\end{table*}

\clearpage

\begin{table*}
\caption{The continuum cores in Sgr B2(M) detected by the QUARKS survey$^{\ (1)}$. \label{coretable}}
\centering
\begin{tabular}{lcccccccc}
\hline\hline
Core ID & RA (J2000) & DEC (J2000) & $L_{\rm max}$ & $L_{\rm min}$& PA$^{\ (2)}$ & $I_{\rm peak}^{(3)}$ & Flux$^{(4)}$ & SM17$^{(5)}$ \\
   &(\degr)& (\degr) & (\arcsec) & (\arcsec)  & (\degr) & (Jy beam$^{-1})$ & Jy & \\
\hline
1 & 266.83403 & -28.38466 & 0.9 & 0.7 & -60.0 & 1.2384 & 2.9881(33) & Y \\
2 & 266.83390 & -28.38451 & 0.7 & 0.5 & -48.0 & 1.0232 & 1.6362(19) & Y \\
3 & 266.83377 & -28.38439 & 0.5 & 0.5 & 0.0 & 0.4462 & 0.5199(7) & N \\
4 & 266.83421 & -28.38456 & 0.6 & 0.5 & 0.0 & 0.4320 & 0.4593(7) & N \\
5 & 266.83379 & -28.38418 & 0.6 & 0.5 & 0.0 & 0.2037 & 0.3029(5) & Y \\
6 & 266.83393 & -28.38489 & 0.7 & 0.6 & 0.0 & 0.3453 & 0.8302(11) & Y \\
7 & 266.83409 & -28.38502 & 0.6 & 0.5 & 0.0 & 0.3427 & 0.4357(5) & Y \\
8 & 266.83362 & -28.38459 & 0.8 & 0.6 & -30.0 & 0.2454 & 0.3665(5) & N \\
9 & 266.83383 & -28.38510 & 0.6 & 0.6 & -51.0 & 0.2598 & 0.4426(7) & Y \\
10 & 266.83416 & -28.38520 & 0.6 & 0.5 & 34.0 & 0.1956 & 0.3105(5) & Y \\
11 & 266.83433 & -28.38535 & 0.6 & 0.5 & 39.0 & 0.1103 & 0.1880(4) & Y \\
12 & 266.83405 & -28.38436 & 0.6 & 0.4 & -6.0 & 0.3207 & 0.3038(5) & N \\
13 & 266.83394 & -28.38397 & 0.7 & 0.5 & 0.0 & 0.0739 & 0.1490(4) & Y \\
14 & 266.83454 & -28.38421 & 0.6 & 0.6 & 0.0 & 0.2274 & 0.2956(5) & Y \\
15 & 266.83449 & -28.38438 & 0.6 & 0.4 & 0.0 & 0.0627 & 0.0839(2) & Y \\
16 & 266.83413 & -28.38421 & 1.0 & 0.6 & 0.0 & 0.1171 & 0.3160(5) & N \\
17 & 266.83379 & -28.38378 & 1.0 & 0.5 & 41.0 & 0.0500 & 0.1045(4) & N \\
18 & 266.83300 & -28.38415 & 0.9 & 0.5 & 90.0 & 0.0638 & 0.1336(4) & Y \\
19 & 266.83359 & -28.38506 & 0.7 & 0.5 & 21.0 & 0.0984 & 0.2074(4) & Y \\
20 & 266.83336 & -28.38496 & 0.7 & 0.5 & 26.0 & 0.0696 & 0.1301(4) & N \\
21 & 266.83314 & -28.38486 & 0.7 & 0.5 & 15.0 & 0.0654 & 0.1284(4) & Y \\
22 & 266.83292 & -28.38476 & 0.8 & 0.5 & 23.0 & 0.0526 & 0.1140(4) & N \\
23 & 266.83264 & -28.38459 & 0.9 & 0.8 & 0.0 & 0.0295 & 0.0910(4) & N \\
24 & 266.83237 & -28.38441 & 0.8 & 0.7 & 0.0 & 0.0178 & 0.0388(2) & N \\
25 & 266.83349 & -28.38527 & 1.1 & 0.8 & -65.0 & 0.0678 & 0.3022(7) & N \\
26 & 266.83289 & -28.38519 & 0.4 & 0.4 & 0.0 & 0.0213 & 0.0187(2) & N \\
27 & 266.83380 & -28.38579 & 0.9 & 0.6 & -69.0 & 0.0520 & 0.1187(4) & Y \\
28 & 266.83304 & -28.38608 & 0.6 & 0.5 & 0.0 & 0.0360 & 0.0467(2) & Y \\
29 & 266.83297 & -28.38591 & 0.6 & 0.6 & 0.0 & 0.0239 & 0.0449(2) & N \\
30 & 266.83306 & -28.38560 & 0.7 & 0.6 & 0.0 & 0.0154 & 0.0369(2) & N \\
31 & 266.83315 & -28.38632 & 1.3 & 1.1 & 90.0 & 0.0190 & 0.1026(5) & N \\
32 & 266.83340 & -28.38643 & 0.6 & 0.5 & -57.0 & 0.0118 & 0.0161(2) & N \\
33 & 266.83272 & -28.38625 & 0.7 & 0.5 & 0.0 & 0.0132 & 0.0193(2) & N \\
34 & 266.83359 & -28.38693 & 1.0 & 0.9 & 90.0 & 0.0171 & 0.0588(4) & N \\
35 & 266.83378 & -28.38710 & 0.7 & 0.6 & 40.0 & 0.0074 & 0.0160(2) & N \\
36 & 266.83372 & -28.38732 & 0.6 & 0.6 & 0.0 & 0.0055 & 0.0081(2) & N \\
37 & 266.83435 & -28.38749 & 0.7 & 0.5 & 0.0 & 0.0046 & 0.0013(2) & N \\
38 & 266.83228 & -28.38615 & 0.8 & 0.6 & 9.0 & 0.0234 & 0.0583(2) & N \\
39 & 266.83207 & -28.38595 & 0.6 & 0.5 & 38.0 & 0.0415 & 0.0597(2) & N \\
40 & 266.83204 & -28.38580 & 0.5 & 0.4 & 90.0 & 0.0262 & 0.0285(2) & Y \\
41 & 266.83175 & -28.38563 & 0.7 & 0.6 & 0.0 & 0.1086 & 0.1925(4) & Y \\
42 & 266.83192 & -28.38617 & 0.6 & 0.4 & 0.0 & 0.0207 & 0.0313(2) & N \\
43 & 266.83193 & -28.38633 & 0.6 & 0.5 & 0.0 & 0.0270 & 0.0367(2) & N \\
44 & 266.83154 & -28.38575 & 0.5 & 0.4 & 0.0 & 0.0245 & 0.0313(2) & N \\
45 & 266.83151 & -28.38605 & 0.8 & 0.7 & 90.0 & 0.0166 & 0.0451(2) & N \\
46 & 266.83192 & -28.38652 & 0.8 & 0.6 & -8.0 & 0.0098 & 0.0205(2) & N \\
47 & 266.83210 & -28.38535 & 0.7 & 0.6 & 0.0 & 0.0284 & 0.0473(2) & Y \\
48 & 266.83159 & -28.38530 & 0.6 & 0.4 & -39.0 & 0.0382 & 0.0480(2) & Y \\
49 & 266.83166 & -28.38539 & 0.4 & 0.3 & 0.0 & 0.0288 & 0.0217(2) & N \\
50 & 266.83135 & -28.38516 & 0.9 & 0.5 & 53.0 & 0.0124 & 0.0259(2) & N \\
51 & 266.83111 & -28.38526 & 0.6 & 0.4 & 0.0 & 0.0162 & 0.0165(2) & N \\
52 & 266.83100 & -28.38547 & 0.7 & 0.4 & 25.0 & 0.0474 & 0.0584(2) & Y \\
53 & 266.83088 & -28.38555 & 0.5 & 0.5 & 0.0 & 0.0251 & 0.0303(2) & N \\
54 & 266.83062 & -28.38583 & 0.9 & 0.6 & 0.0 & 0.0226 & 0.0458(2) & Y \\
55 & 266.83109 & -28.38703 & 0.6 & 0.6 & 0.0 & 0.0081 & 0.0059(2) & N \\
\hline
\end{tabular}{\flushleft
$^{(1)}$ This table contains 97 rows.\\
$^{(2)}$ The position angle (PA) is defined in the style of DS9. When PA is 0\degr, the major axis
is in the direction of east. The major axis rotates  from the east to the north when PA changes from 0\degr~to 
90 \degr.  \\
$^{(3)}$ The uncertainty of the $I_{\rm peak}$ is estimated to be 1 mJy beam$^{-1}$ for weak cores. \\
$^{(4)}$ The number in the bracket is the uncertainty of the last digital of the flux.\\
$^{(5)}$ Denote if the corresponding core has been identified by \citet{2017A&A...604A...6S}. \\
}
\end{table*}

\addtocounter{table}{-1}

\begin{table*}
\caption{continued.}
\centering
\begin{tabular}{ccccccccc}
\hline\hline
Core ID & RA (J2000) & DEC (J2000) & $L_{\rm max}$ & $L_{\rm min}$& PA & $I_{\rm peak}$ & Flux & SM17 \\
   &(\degr)& (\degr) & (\arcsec) & (\arcsec)  & (\degr) & (Jy beam$^{-1})$ & Jy & \\
\hline
56 & 266.83333 & -28.38848 & 1.1 & 1.1 & 90.0 & 0.0100 & 0.0320(5) & N \\
57 & 266.83148 & -28.38423 & 0.6 & 0.5 & 0.0 & 0.0453 & 0.0664(2) & Y \\
58 & 266.83151 & -28.38408 & 0.4 & 0.4 & 90.0 & 0.0274 & 0.0228(2) & N \\
59 & 266.83118 & -28.38441 & 0.5 & 0.5 & 0.0 & 0.0154 & 0.0154(2) & N \\
60 & 266.83079 & -28.38422 & 0.6 & 0.4 & 30.0 & 0.0231 & 0.0289(2) & Y \\
61 & 266.83172 & -28.38373 & 1.1 & 0.6 & 0.0 & 0.0063 & 0.0150(2) & N \\
62 & 266.83121 & -28.38339 & 0.5 & 0.5 & 0.0 & 0.0165 & 0.0174(2) & N \\
63 & 266.83107 & -28.38326 & 0.7 & 0.7 & 0.0 & 0.0221 & 0.0492(2) & Y \\
64 & 266.83085 & -28.38320 & 0.7 & 0.5 & 0.0 & 0.0137 & 0.0212(2) & N \\
65 & 266.83039 & -28.38351 & 1.6 & 1.1 & -51.0 & 0.0067 & 0.0416(5) & N \\
66 & 266.83007 & -28.38412 & 0.5 & 0.4 & 0.0 & 0.0128 & 0.0106(2) & N \\
67 & 266.82988 & -28.38406 & 0.4 & 0.4 & -55.0 & 0.0089 & 0.0079(2) & N \\
68 & 266.82947 & -28.38455 & 0.8 & 0.7 & -31.0 & 0.0281 & 0.0585(2) & N \\
69 & 266.82931 & -28.38468 & 0.5 & 0.5 & 0.0 & 0.0082 & 0.0077(2) & N \\
70 & 266.82923 & -28.38447 & 0.7 & 0.6 & 0.0 & 0.0102 & 0.0193(2) & N \\
71 & 266.82954 & -28.38401 & 0.9 & 0.7 & 90.0 & 0.0096 & 0.0296(2) & N \\
72 & 266.82939 & -28.38363 & 0.5 & 0.4 & 0.0 & 0.0072 & 0.0063(2) & N \\
73 & 266.82919 & -28.38513 & 0.6 & 0.5 & 72.0 & 0.0087 & 0.0122(2) & N \\
74 & 266.82901 & -28.38517 & 0.5 & 0.4 & 0.0 & 0.0126 & 0.0114(2) & N \\
75 & 266.82871 & -28.38522 & 0.8 & 0.4 & -25.0 & 0.0107 & 0.0172(2) & N \\
76 & 266.82813 & -28.38442 & 0.6 & 0.6 & 0.0 & 0.0399 & 0.0576(2) & N \\
77 & 266.82837 & -28.38439 & 0.5 & 0.5 & 0.0 & 0.0165 & 0.0187(2) & N \\
78 & 266.82810 & -28.38234 & 1.2 & 0.8 & 0.0 & 0.0306 & 0.1156(5) & N \\
79 & 266.83354 & -28.38364 & 0.8 & 0.5 & 0.0 & 0.0173 & 0.0367(2) & N \\
80 & 266.83329 & -28.38363 & 0.7 & 0.7 & 0.0 & 0.0099 & 0.0236(2) & N \\
81 & 266.83333 & -28.38305 & 1.1 & 0.9 & 0.0 & 0.0056 & 0.0184(4) & N \\
82 & 266.83360 & -28.38299 & 1.1 & 0.7 & 90.0 & 0.0055 & 0.0136(4) & N \\
83 & 266.83382 & -28.38290 & 0.8 & 0.7 & 0.0 & 0.0057 & 0.0132(2) & N \\
84 & 266.83417 & -28.38275 & 1.0 & 0.9 & 90.0 & 0.0074 & 0.0256(4) & N \\
85 & 266.83135 & -28.38218 & 1.1 & 0.9 & 90.0 & 0.0054 & 0.0133(4) & N \\
86 & 266.82855 & -28.38578 & 0.7 & 0.6 & 0.0 & 0.0115 & 0.0181(2) & N \\
87 & 266.83417 & -28.38583 & 0.9 & 0.8 & 90.0 & 0.0169 & 0.0597(4) & N \\
88 & 266.83468 & -28.38563 & 0.5 & 0.5 & 90.0 & 0.0233 & 0.0276(2) & N \\
89 & 266.83469 & -28.38546 & 0.6 & 0.5 & 15.0 & 0.0254 & 0.0431(2) & N \\
90 & 266.83490 & -28.38548 & 0.8 & 0.7 & 0.0 & 0.0219 & 0.0679(4) & Y \\
91 & 266.83485 & -28.38524 & 0.9 & 0.8 & 90.0 & 0.0239 & 0.0998(4) & N \\
92 & 266.83496 & -28.38479 & 1.5 & 1.0 & -50.0 & 0.0580 & 0.3959(9) & Y \\
93 & 266.83539 & -28.38452 & 1.8 & 0.9 & 0.0 & 0.0358 & 0.2894(8) & N \\
94 & 266.83504 & -28.38655 & 1.9 & 1.8 & 0.0 & 0.0067 & 0.0579(8) & N \\
95 & 266.83296 & -28.38263 & 1.5 & 1.1 & -56.0 & 0.0037 & 0.0067(5) & N \\
96 & 266.83133 & -28.38798 & 0.6 & 0.5 & -30.0 & 0.0052 & 0.0062(2) & N \\
97 & 266.83254 & -28.38526 & 0.5 & 0.5 & 0.0 & 0.0043 & 0.0053(2) & N \\
\hline
\end{tabular}
\end{table*}

\end{CJK*}
\end{document}